\documentclass[twocolumn]{aastex63}

\usepackage[utf8]{inputenc}
\usepackage{graphicx,ulem}
\usepackage{longtable}
\usepackage{booktabs}
\usepackage{float}
\usepackage{dcolumn}
\usepackage{graphics,epsfig}
\usepackage{amsmath,amssymb,latexsym,mathrsfs}
\usepackage{bm}
\usepackage{color}
\usepackage{subfigure}
\usepackage{soul}
\usepackage{multirow}   

\newcommand{\de}{\text{d}}
\newcommand{\sz}{{\rm SZ}}
\newcommand{\mwl}{M_{\rm WL}}

\received{June 1, 2019}
\revised{January 10, 2019}
\accepted{\today}
\submitjournal{APJ}

\newcommand{\McGill}{Department of Physics and McGill Space Institute, McGill University, Montreal, Quebec H3A 2T8, Canada}
\newcommand{\KIPAC}{Kavli Institute for Particle Astrophysics and Cosmology, Stanford University, 452 Lomita Mall, Stanford, CA 94305}
\newcommand{\Stanford}{Dept. of Physics, Stanford University, 382 Via Pueblo Mall, Stanford, CA 94305}

\newcommand{\KICPChicago}{Kavli Institute for Cosmological Physics, University of Chicago, Chicago, IL, USA 60637}
\newcommand{\PhysicsUChicago}{Department of Physics, University of Chicago, Chicago, IL, USA 60637}
\newcommand{\AAUChicago}{Department of Astronomy and Astrophysics, University of Chicago, Chicago, IL, USA 60637}
\newcommand{\FNAL}{Fermi National Accelerator Laboratory, MS209, P.O. Box 500, Batavia, IL 60510}
\newcommand{\ArgonneHEP}{High Energy Physics Division, Argonne National Laboratory, Argonne, IL, USA 60439}
\newcommand{\EFIChicago}{Enrico Fermi Institute, University of Chicago, Chicago, IL, USA 60637}

\newcommand{\Caltech}{California Institute of Technology, Pasadena, CA, USA 91125}
\newcommand{\Berkeley}{Department of Physics, University of California, Berkeley, CA, USA 94720}
\newcommand{\Cifar}{Canadian Institute for Advanced Research, CIFAR Program in Cosmology and Gravity, Toronto, ON, M5G 1Z8, Canada}
\newcommand{\Colorado}{Center for Astrophysics and Space Astronomy, Department of Astrophysical and Planetary Sciences, University of Colorado, Boulder, CO, 80309}
\newcommand{\ESO}{European Southern Observatory, Karl-Schwarzschild-Stra{\ss}e 2, 85748 Garching, Germany}
\newcommand{\Colphys}{Department of Physics, University of Colorado, Boulder, CO, 80309}
\newcommand{\Illast}{Astronomy Department, University of Illinois at Urbana-Champaign, 1002 W. Green Street, Urbana, IL 61801, USA}
\newcommand{\Illphys}{Department of Physics, University of Illinois Urbana-Champaign, 1110 W. Green Street, Urbana, IL 61801, USA}
\newcommand{\UChicago}{University of Chicago, Chicago, IL, USA 60637}
\newcommand{\LBNL}{Physics Division, Lawrence Berkeley National Laboratory, Berkeley, CA, USA 94720}
\newcommand{\Mcmaster}{Department of Physics and Astronomy, McMaster University, 1280 Main St. W., Hamilton, ON L8S 4L8, Canada}

\newcommand{\Munich}{Faculty of Physics, Ludwig-Maximilians-Universit\"{a}t, 81679 M\"{u}nchen, Germany}
\newcommand{\ExcellenceCluster}{Excellence Cluster ORIGINS, Boltzmannstr. 2, 85748 Garching, Germany}
\newcommand{\MPE}{Max-Planck-Institut f\"{u}r extraterrestrische Physik, 85748 Garching, Germany}

\newcommand{\Minnesota}{Department of Physics, University of Minnesota, Minneapolis, MN, USA 55455}
\newcommand{\Melbourne}{School of Physics, University of Melbourne, Parkville, VIC 3010, Australia}
\newcommand{\CaseWestern}{Physics Department, Center for Education and Research in Cosmology and Astrophysics, Case Western Reserve University,Cleveland, OH, USA 44106}
\newcommand{\ArtInstChicago}{Liberal Arts Department, School of the Art Institute of Chicago, Chicago, IL, USA 60603}
\newcommand{\JPL}{Jet Propulsion Laboratory, California Institute of Technology, Pasadena, CA 91109, USA}
\newcommand{\CfA}{Center for Astrophysics $|$ Harvard \& Smithsonian, 60 Garden Street, Cambridge, MA 02138, USA}

\newcommand{\Toronto}{Department of Astronomy \& Astrophysics, University of Toronto, 50 St George St, Toronto, ON, M5S 3H4, Canada}

\newcommand{\KEK}{High Energy Accelerator Research Organization (KEK), Tsukuba, Ibaraki 305-0801, Japan}

\shorttitle{SPT-Planck}
\shortauthors{Salvati et al.}

\begin{document}

\title{Combining Planck and SPT cluster catalogs: cosmological analysis and impact on Planck scaling relation calibration
\footnote{Released on ...}}

\correspondingauthor{Laura Salvati}
\email{laura.salvati@inaf.it,\\
laura.salvati@universite-paris-saclay.fr}

\author{L.~Salvati}
\affiliation{INAF – Osservatorio Astronomico di Trieste, Via G. B. Tiepolo 11, 34143 Trieste, Italy}
\affiliation{IFPU – Institute for Fundamental Physics of the Universe, Via Beirut 2, 34014 Trieste, Italy}
\affiliation{Université Paris-Saclay, CNRS, Institut d'Astrophysique Spatiale, 91405, Orsay, France}

\author{A.~Saro}
\affiliation{Dipartimento di Fisica, Sezione di Astronomia, Università di Trieste, Via Tiepolo 11, I-34143 Trieste, Italy}
\affiliation{INAF – Osservatorio Astronomico di Trieste, Via G. B. Tiepolo 11, 34143 Trieste, Italy}
\affiliation{IFPU – Institute for Fundamental Physics of the Universe, Via Beirut 2, 34014 Trieste, Italy}
\affiliation{INFN – Sezione di Trieste, I-34100 Trieste, Italy}

\author{S.~Bocquet}
\affiliation{Max-Planck-Institut f\"{u}r  Astrophysik (MPA), Karl-Schwarzschild Strasse 1, 85748 Garching bei M\"{u}nchen, Germany}

\author{M.~Costanzi}
\affiliation{INAF – Osservatorio Astronomico di Trieste, Via G. B. Tiepolo 11, 34143 Trieste, Italy}
\affiliation{IFPU – Institute for Fundamental Physics of the Universe, Via Beirut 2, 34014 Trieste, Italy}
\affiliation{Dipartimento di Fisica, Sezione di Astronomia, Università di Trieste, Via Tiepolo 11, I-34143 Trieste, Italy}

\author{B.~Ansarinejad}
\affiliation{\Melbourne}

\author[0000-0002-5108-6823]{B.~A.~Benson}
\affiliation{\FNAL}
\affiliation{\KICPChicago}
\affiliation{\AAUChicago}

\author[0000-0001-7665-5079]{L.~E.~Bleem}
\affiliation{High Energy Physics Division, Argonne National Laboratory, 9700 South Cass Avenue, Lemont, IL 60439, USA}
\affiliation{Kavli Institute for Cosmological Physics, University of Chicago, 5640 South Ellis Avenue, Chicago, IL 60637, USA}

\author[0000-0002-2238-2105]{M.~S.~Calzadilla}
\affiliation{Kavli Institute for Astrophysics and Space Research, Massachusetts Institute of Technology Cambridge, MA 02139, USA}

\author[0000-0002-2044-7665]{J.~E.~Carlstrom}
\affiliation{\KICPChicago}
\affiliation{\AAUChicago}
\affiliation{\PhysicsUChicago}
\affiliation{\ArgonneHEP}
\affiliation{\EFIChicago}

\author[0000-0002-6311-0448]{C.~L.~Chang}
\affiliation{\ArgonneHEP}
\affiliation{\KICPChicago}
\affiliation{\AAUChicago}

\author[0000-0001-8241-7704]{R.~Chown}
\affiliation{\Mcmaster}

\author{A.~T.~Crites}
\affiliation{\Toronto}
\affiliation{\KICPChicago}
\affiliation{\AAUChicago}

\author{T.~de~Haan}
\affiliation{\KEK}
\affiliation{\Berkeley}

\author{M.~A.~Dobbs}
\affiliation{\McGill}
\affiliation{\Cifar}

\author[0000-0002-5370-6651] {W.~B.~Everett}
\affiliation{\Colorado}

\author{B.~Floyd}
\affiliation{Department of Physics and Astronomy, University of Missouri--Kansas City, 5110 Rockhill Road, Kansas City, MO 64110, USA}

\author{S.~Grandis}
\affiliation{Faculty of Physics, Ludwig-Maximilians-Universität, Scheinerstr. 1, 81679, Munich, Germany}

\author[0000-0001-7874-0445]{E.~M.~George}
\affiliation{\ESO}
\affiliation{\Berkeley}

\author[0000-0003-2606-9340]{N.~W.~Halverson}
\affiliation{\Colorado}
\affiliation{\Colphys}

\author{G.~P.~Holder}
\affiliation{\Illast}
\affiliation{\Illphys}

\author{W.~L.~Holzapfel}
\affiliation{\Berkeley}

\author{J.~D.~Hrubes}
\affiliation{\UChicago}

\author[0000-0003-3106-3218]{A.~T.~Lee}
\affiliation{\Berkeley}
\affiliation{\LBNL}

\author{D.~Luong-Van}
\affiliation{\UChicago}

\author{M.~McDonald}
\affiliation{Kavli Institute for Astrophysics and Space Research, Massachusetts Institute of Technology, 77 Massachusetts Avenue, Cambridge, MA 02139}

\author{J.~J.~McMahon}
\affiliation{\KICPChicago}
\affiliation{\AAUChicago}
\affiliation{\PhysicsUChicago}
\affiliation{\EFIChicago}

\author{S.~S.~Meyer}
\affiliation{\KICPChicago}
\affiliation{\AAUChicago}
\affiliation{\PhysicsUChicago}
\affiliation{\EFIChicago}

\author{M.~Millea}
\affiliation{\Berkeley}

\author{L.~M.~Mocanu}
\affiliation{\KICPChicago}
\affiliation{\AAUChicago}

\author[0000-0002-6875-2087]{J.~J.~Mohr}
\affiliation{\Munich}
\affiliation{\ExcellenceCluster}
\affiliation{\MPE}

\author{T.~Natoli}
\affiliation{\KICPChicago}
\affiliation{\AAUChicago}

\author{Y.~Omori}
\affiliation{\KICPChicago}
\affiliation{\AAUChicago}
\affiliation{\KIPAC}
\affiliation{\Stanford}

\author{S.~Padin}
\affiliation{\Caltech}
\affiliation{\KICPChicago}
\affiliation{\AAUChicago}

\author{C.~Pryke}
\affiliation{\Minnesota}

\author[0000-0003-2226-9169]{C.~L.~Reichardt}
\affiliation{\Melbourne}

\author{J.~E.~Ruhl}
\affiliation{\CaseWestern}

\author{F.~Ruppin}
\affiliation{Univ. Lyon, Univ. Claude Bernard Lyon 1, CNRS/IN2P3, IP2I Lyon, F‐69622, Villeurbanne, France}
\affiliation{Kavli Institute for Astrophysics \& Space Research, Massachusetts Institute of Technology, 77 Massachusetts Ave., Cambridge, MA 02139, USA}

\author{K.~K.~Schaffer}
\affiliation{\ArtInstChicago}
\affiliation{\KICPChicago}
\affiliation{\EFIChicago}

\author[0000-0002-6987-7834]{T.~Schrabback}
\affiliation{Argelander Institut f\"ur Astronomie, Auf dem H\"ugel 71, D-53121 Bonn, Germany}

\author[0000-0002-2757-1423]{E.~Shirokoff}
\affiliation{\KICPChicago} 
\affiliation{\AAUChicago} 
\affiliation{\Berkeley} 

\author{Z.~Staniszewski}
\affiliation{\JPL}
\affiliation{\CaseWestern}

\author[0000-0002-2718-9996]{A.~A.~Stark}
\affiliation{\CfA}

\author[0000-0001-7192-3871]{J.~D.~Vieira}
\affiliation{\Illast} 
\affiliation{\Illphys} 

\author{R.~Williamson}
\affiliation{\JPL}
\affiliation{\KICPChicago} 
\affiliation{\AAUChicago}

\begin{abstract}
We provide the first combined cosmological analysis of South Pole Telescope (SPT) and Planck cluster catalogs. The aim is to provide an independent calibration for Planck scaling relations, exploiting the cosmological constraining power of the SPT-SZ cluster catalog and its dedicated weak lensing (WL) and X-ray follow-up observations. We build a new version of the Planck cluster likelihood. In the $\nu \Lambda$CDM scenario, focusing on the mass slope and mass bias of Planck scaling relations, we find $\alpha_{\text{SZ}} = 1.49 _{-0.10}^{+0.07}$ and $(1-b)_{\text{SZ}} = 0.69 _{-0.14}^{+0.07}$ respectively. The results for the mass slope show a $\sim 4 \, \sigma$ departure from the self-similar evolution, $\alpha_{\text{SZ}} \sim 1.8$. This shift is mainly driven by the matter density value preferred by SPT data, $\Omega_m = 0.30 \pm 0.03$, lower than  the one obtained by Planck data alone, $\Omega_m = 0.37 _{-0.06}^{+0.02}$.
The mass bias constraints are consistent both with outcomes of hydrodynamical simulations and external WL calibrations, $(1-b) \sim 0.8$, and with results required by the Planck cosmic microwave background cosmology, $(1-b) \sim 0.6$. 
From this analysis, we obtain a new catalog of Planck cluster masses $M_{500}$. 
We estimate the ratio between the published Planck $M_{\text{SZ}}$ masses and our derived masses $M_{500}$, as a ``measured mass bias", $(1-b)_M$.
We analyse the mass, redshift and detection noise dependence of $(1-b)_M$, finding an increasing trend towards high redshift and low mass. 
These results mimic the effect of departure from self-similarity in cluster evolution, showing different dependencies for the low-mass high-mass, low-z high-z regimes.
\end{abstract}


\section{Introduction} \label{sec:intro}

Galaxy clusters are the largest, gravitationally bound structures in the Universe. These objects represent the nodes in the cosmic web of the large scale structure, and are related to the peaks in the density field, on scales of the order of megaparsec.

Galaxy clusters can be detected in different wavelengths. In recent years, several experiments produced large catalogs of clusters to be used for the cosmological analysis, such as the Planck survey \citep{2016A&A...594A..27P,2016A&A...594A..24P}, the South Pole Telescope (SPT hereafter) \citep{Bleem:2014iim,2016ApJ...832...95D,Bocquet:2018ukq} and the Atacama Cosmology Telescope \citep{Hilton:2020qsa} in the millimeter wavelengths; the Kilo-Degree Survey\citep{2019MNRAS.485..498M}, the Dark Energy Survey \citep{DES:2017myt,Abbott:2020knk} in optical; the ROSAT survey \citep{Boehringer:2017wvr}, the XXL survey \citep{2018A&A...620A...5A,2018A&A...620A..10P} and the first eROSITA observations \citep{Liu:2021ewv} in X-rays. 
In particular, the abundance of galaxy clusters (galaxy cluster number counts) has emerged as a fundamental cosmological probe. Cluster formation and evolution is strictly related to the underlying cosmological model, tracing the growth of structures, see e.g. \cite{Allen:2011zs}. In particular, the observed cluster abundance is mainly sensitive to the combination of two cosmological parameters: the total matter density $\Omega_m$ and $\sigma_8$, which is defined as the rms fluctuation in the linear matter density field on $8 \, \text{Mpc}/h$ scale at redshift $z=0$.
Comparing and combining results from cluster abundance with
other cosmological probes, such as cosmic microwave background radiation (CMB hereafter) at high redshift, or baryon acoustic oscillations (BAO hereafter) at low redshift, allows us to perform fundamental consistency checks of the standard cosmological model.

Cosmological constraints from cluster counts rely
on the knowledge of their mass and redshift distribution, which is described by the halo mass function, see e.g. discussion in \cite{Monaco:2016pys} and references therein for an updated list of available mass function evaluations, and \cite{2019ApJ...872...53M,2020ApJ...901....5B} for recent mass function emulators.
However, cluster mass cannot be measured directly
forcing us to rely on observational mass-proxies that correlate with the underlying halo mass.

Cluster masses and survey observables are linked through statistical scaling relations, that describe the interplay between astrophysics and cosmology in the cluster formation and evolution.
These relations are usually calibrated through a multi-wavelength analysis. Indeed, observations of the same clusters in different frequency bands provide a unique insight on the interaction between baryonic and dark matter, allowing us to further model the impact of astrophysical processes on the cluster cosmological evolution.
Scaling relations are then combined with a model for the selection process (i.e. a selection function) to transform the theoretical halo mass function into a prediction for the distribution of clusters in the space of redshift and survey observables. In this scenario, it is clear that a precise and comprehensive characterization of the mass function, the scaling relations and the selection function is needed in order to provide stringent and unbiased constraints on cosmological parameters from galaxy clusters.

In this work, we perform the first combined cosmological analysis of the SPT-SZ \citep{Bleem:2014iim} and Planck \citep{2016A&A...594A..27P} cluster catalogs. Both experiments detect clusters in the millimeter wavelengths, through the thermal Sunyaev-Zeldovich (tSZ hereafter) effect \citep{Sunyaev:1970er}.
The strength of this analysis lies in the combination of a full-sky survey (Planck) with deep and high-resolution observations from a ground-based experiment (SPT). 
The combination of the two cluster catalogs spans a large redshift range (from $z=0$ for Planck catalog, up to $z\sim 1.7$ for the SPT one), ensuring the possibility to test 
the impact of astrophysics over a broad redshift range.
The strength of combining Planck and SPT cluster observations has been already explored in the analysis of \cite{Melin:2020wuy}, in which the authors provide a new cluster catalog extracted from the common area observed by the two experiments.
The analysis we present here is the first in a series of papers in which we plan to exploit the combination of the SPT-SZ and Planck
cluster catalogs. In this work we focus primarily on providing a new calibration for Planck scaling relations.

In \cite{2016A&A...594A..24P} the evaluation of Planck cluster masses from tSZ observations is based on the assumption of hydrostatic equilibrium (HE hereafter). Hydrodynamical simulations suggest, however, that HE cluster masses are biased low by a factor of $\sim 20\%$, see e.g., discussion in \citep{Pratt:2019cnf}. A mass-bias parameter, defined through the ratio between the HE inferred mass and the total cluster mass, is thus introduced: $(1-b) = M_{\rm{SZ}}/M_{\rm{tot}} \sim 0.8$.
The calibration of the whole mass-observable scaling relation is done through external X-ray and weak lensing (WL hereafter) measurements, the latter used in particular to estimate the mass bias. Nevertheless, WL analyses based on different cluster subsamples and approaches \citep{vonderLinden:2014haa,Hoekstra:2015gda,Okabe:2015faa,Sereno:2015jda,Smith:2015qhs,Penna-Lima:2016tvo,Sereno:2017zcn,Herbonnet:2019byy} might provide different mass calibrations, resulting in different constraints on cosmological parameters and showing therefore the impact of the cluster subsample selection choice, see also discussion in \citealt{Salvati:2019zdp}.

The mass calibration plays therefore an important role in the CMB-cluster $\sigma_8$ tension \citep{2014A&A...571A..20P,2016A&A...594A..24P}, where the discrepancy could be entirely relieved by adopting a mass-bias parameter of  $(1-b) \sim 0.6$. Such a strong deviation from HE masses would be, however, in strong contrast with the above described WL observations and hydrodynamical simulations predictions, and with several other astrophysical observations for clusters \citep[see, e.g.,][]{Eckert:2018mlz}. Finally, we note that more recent analyses of Planck data \citep{Aghanim:2016yuo,Salvati:2017rsn,2020A&A...641A...6P}
reveal that cosmological results are now consistent between CMB primary anisotropies and galaxy clusters, with constraints on the $\sigma_8$ parameter well in agreement within $2 \, \sigma$. These results are still systematically limited by the assumed mass calibration, which, as in the original Planck analysis, strongly depends on the subsample of clusters adopted to constrain the mass-bias parameter.

It is therefore fundamental to perform an independent calibration of the scaling relations. The Planck and SPT-SZ cluster catalog combination that we propose in this work specifically address this point. By exploiting the cosmological constraining power of SPT-SZ clusters and its associated mass-calibration data-sets, and the tight correlation between cosmology and astrophysics, we provide an independent evaluation of Planck scaling relation parameters and a new evaluation of Planck total cluster masses, which are therefore consistent with the SPT-SZ WL calibrated masses and corrected for Eddington bias effects \citep{Allen:2011zs}.

The paper is structured as follows: in section \ref{sec:data} we describe the cluster observations for Planck and SPT and 
the underlying theoretical model for the use of cluster number counts. In section \ref{sec:method} we discuss the approach used to combine the datasets and extract cosmological information, and the recipe to evaluate Planck cluster masses and further analyse the mass bias. We present and discuss the results in sections \ref{sec:results} and \ref{sec:discussion}, and derive our final conclusions in section \ref{sec:conclusion}.

\section{Data and Model} \label{sec:data}

In this section we summarise the observation and detection strategies for the Planck and SPT experiments. We also describe the theoretical models that lead to the evaluation of the likelihood function needed for the cosmological analysis. For the full discussion, we refer to 
the SPT analysis in \cite{Bleem:2014iim,Bocquet:2018ukq} and the Planck analysis in \cite{2016A&A...594A..24P,2016A&A...594A..27P}.

We recall here that clusters detected through the tSZ effect are often defined as objects with a mass $M_{500}$ contained in a sphere of radius $R_{500}$, such that the cluster mean mass overdensity inside $R_{500}$ corresponds to 500 times the critical density $\rho_c(z)$. Therefore we define the total cluster mass as
\begin{equation}
M_{500} = \frac{4\pi}{3}R^3_{500} 500 \rho_c(z) \, .    
\end{equation}

\subsection{South Pole Telescope}\label{sec:SPT}

The South Pole Telescope is a 10m diameter telescope located at the geographic South Pole \citep{Carlstrom:2009um}. 
We consider observations of the SPT-SZ survey \citep{Bleem:2014iim}, which detected galaxy clusters through the tSZ effect, using observations in the $95$ and $150$ GHz bands, in a $2500$ deg$^2$ area.
With $\sim 1'$ resolution and $1^{\circ}$ field of view, SPT is able to observe rare, high-mass clusters, from redshift $z\gtrsim 0.2$.

Galaxy clusters are extracted from the SPT-SZ survey data through a multi-matched filter technique, see e.g \cite{Melin:2006qq}. This approach makes use of the known (non-relativistic) tSZ spectral signature and a model for the spatial profile of the signal. 
In the standard SPT analysis approach,
the spatial profile follows the projected isothermal $\beta$ model \citep{Cavaliere:1976tx}, with $\beta$ fixed to 1.

The tSZ signature is then used, together with a description of the noise sources in the frequency maps, to construct a filter designed to maximize the sensitivity to galaxy clusters.
From the filtered maps, we can extract cluster candidates, via a peak detection algorithm similar to the \texttt{SExtractor} routine \citep{Bertin:1996fj}. In SPT analysis, the maximum detection significance (the signal-to-noise ratio maximized over all filter scales) $\xi$ is used as tSZ observable.

In this work we focus on the cosmological cluster sample, analyzed in \cite{2016ApJ...832...95D} and \cite{Bocquet:2018ukq}. It is a subsample of the full SPT-SZ sample, consisting of 365 detections (343 of which have been optically confirmed), restricted to $z>0.25$ and with a detection significance $\xi > 5$.

For the SPT cluster cosmological analysis, we follow the recipe described in \cite{Bocquet:2018ukq}. We report here the main steps and refer the reader to the original study for further details.
We make use of a multi-wavelength approach, considering also WL and X-ray data. In detail, we use WL measurements of 32 clusters in the SPT-SZ cosmological sample, considering the reduced tangential shear profiles in angular coordinates (corrected for contamination by cluster galaxies) and the estimated redshift distributions of the selected source galaxies. These measurements are obtained with Magellan/Megacam \citep{Dietrich:2017zmi} for 19 clusters in the redshift range $0.29\leq z \leq 0.69$, and with the Advanced Camera for Surveys on board of the Hubble Space Telescope (HST hereafter) \citep{Schrabback:2016hac} for 13 clusters in the redshift range $0.576 \leq z \leq 1.132$. 
For the X-ray measurements, we consider 
Chandra observations for 89 clusters in the SPT-SZ cosmological sample \citep{McDonald:2013fka,McDonald:2017ypo}.
The X-ray data products used in this analysis are the total gas mass $M_{\rm gas}$ within an outer radius ranging from $80$ to $2000$ kpc, and the spectroscopic temperature $T_X$ in the $0.15 R_{500}-R_{500}$ range. 

The SPT cluster cosmological analysis is based on a multi-observable Poissonian likelihood. 
The likelihood function can be written as
\begin{eqnarray}\label{eq:spt_likelihood1}
\ln {\mathscr{L}_{\text{SPT}}} &=& \sum _i \ln{\dfrac{dN(\xi,z|\bf{p})}{d\xi dz}} \vert _{\xi_i,z_i} \notag \\
& - & \int _{z_{\rm cut}} ^{\infty} dz \int _{\xi_{\rm cut}}^{\infty} d\xi \dfrac{dN(\xi,z|\bf{p})}{d\xi dz} \notag \\
& + & \sum_j \ln{P(Y_X,g_t|\xi_j,z_j,\bf{p})} \vert _{Y_{X_j},g_{t_j}} \, .
\end{eqnarray}
In the above equation, $\bf{p}$ is the vector of cosmological and scaling relation parameters, the first sum is over all the $i$ clusters in the cosmological sample, while the second sum is for the $j$ clusters with $Y_X=M_{\text{gas}}T_X$ and/or WL measurements, with $g_t$ being the reduced tangential shear profile. Therefore, the first two terms represent the tSZ cluster abundance, while the third 
encodes the information from follow-up mass calibration data.

In order to account for the impact of noise bias on the detection significance $\xi$, we introduce the unbiased tSZ significance $\zeta$. It is defined as the signal-to-noise ratio at the true, underlying cluster position and filter scale. The relation between the two quantities, across many noise realizations, is given by
\begin{equation}\label{eq:zeta_xi}
P(\xi|\zeta) = \mathcal{N}(\sqrt{\zeta^2+3},1)\, .
\end{equation}
This definition has been largely tested and validated in \cite{2010ApJ...722.1180V} and \cite{,2016ApJ...832...95D}.
We can now explicitly evaluate the different terms in Eq.~\ref{eq:spt_likelihood1}. The first term is given by
\begin{eqnarray}\label{eq:spt_likelihood2}
\dfrac{dN(\xi,z|\bf{p})}{d\xi dz} &=& \iint dM_{500} \, d\zeta  \,\, [ P(\xi|\zeta) P(\zeta|M_{500},z,\bf{p})  \notag \\
& \times & \dfrac{dN(M_{500},z|\bf{p})}{dM_{500}} \, \Omega (z,\bf{p}) ] \, .
\end{eqnarray}
In the above equation, $\Omega (z,\bf{p})$ is the survey volume, $dN(M_{500},z|\textbf{p})/dM_{500}$ is the halo mass function, $P(\zeta|M_{500},z,\bf{p})$ 
is the unbiased observable-mass relation
and $P(\xi|\zeta)$ is the measurement uncertainty defined in Eq.~\ref{eq:zeta_xi}.
Therefore, the first term in Eq.~\ref{eq:spt_likelihood1}
is obtained evaluating Eq.~\ref{eq:spt_likelihood2} at the measured $(\xi_i,z_i)$ for each cluster, marginalizing over photometric redshift errors where present.
The second term is 
simply evaluated through a two-dimensional integral over Eq.~\ref{eq:spt_likelihood2}.

The last term in Eq.~\ref{eq:spt_likelihood1} represents the mass calibration contribution and can be evaluated as

\begin{align}\label{eq:spt_likelihood3}
& P(Y_X^{\text{obs}},g_t^{\text{obs}}|\xi,z,\bf{p}) \nonumber \\
& = \iiiint dM_{500} d\zeta dY_X dM_{\text{WL}} \nonumber \\
& \times \left[ P(Y_X^{\text{obs}}|Y_X) P(g_t^{\text{obs}}|M_{\text{WL}}) P(\xi|\zeta)  \right. \nonumber \\
& \times \left. P(\zeta,Y_X,M_{\text{WL}}|M_{500},z,\textbf{p}) P(M_{500}|z,\bf{p})\right] \, ,
\end{align}
where $P(M_{500}|z,\textbf{p})$ is the normalized halo mass function. 
The multi-observable scaling relation $P(\zeta,Y_X,M_{\text{WL}}|M_{500},z,\textbf{p})$ is assumed to follow a multivariate lognormal distribution, whose mean values, for the unbiased tSZ significance $\zeta$, 
the X-ray $Y_X$ quantity and the WL mass $M_{\rm WL}$, read:
\begin{eqnarray}\label{eq:spt_SR1}
\langle \ln \zeta \rangle &=& \ln{A_\sz} + B_\sz \left( \dfrac{M_{500}h_{70}}{4.3 \times 10^{14}\odot} \right) \notag \\
& + & C_\sz \ln{\left( \dfrac{E(z)}{E(0.6)} \right)}
\end{eqnarray}

\begin{eqnarray}\label{eq:spt_SR2}
\ln{\left( \dfrac{M_{500}h_{70}}{8.37 \times 10^{13}M_{\odot}} \right)} &=& \ln{A_{Y_X}} + B_{Y_X} \langle \ln{Y_x} \rangle \notag \\
&+& B_{Y_x} \ln {\left( \dfrac{h_{70}^{5/2}}{3\times10^{14} M_{\odot} \rm keV} \right)} \notag \\
&+& C_{Y_X} \ln{E(z)}
\end{eqnarray}

\begin{equation}\label{eq:spt_SR3}
\langle \ln M_{\rm WL}\rangle = \ln{b_{\rm WL}} + \ln{M_{500}} \, . 
\end{equation}

\noindent The covariance matrix elements of the multi-observable scaling relation are defined as $C_{ij}=\rho( \mathcal{O}_i;\mathcal{O}_j) \sigma_\mathcal{O}^i \sigma_\mathcal{O}^j$, where the intrinsic scatters $\sigma_\mathcal{O}$ of the observables $\mathcal{O}={\zeta,Y_x,M_{\rm WL}}$ are assumed to be independent of mass and redshift, and the three coefficients $\rho( \mathcal{O}_i;\mathcal{O}_j)$ account for their correlations. The full description of the WL bias, $b_{\text{WL}}$, and the associated scatter is done in \cite{Bocquet:2018ukq}, we only recall here that the modelling introduces six nuisance parameters $\delta_i$. All the parameters characterizing the scaling relations are listed and defined in Table~\ref{tab:parameters_ALL}.

We conclude mentioning that the SPT-SZ cosmological sample contains 22 tSZ detections with unknown redshift, since they have not been confirmed through optical counterparts. 
This number is consistent with the expected number of false detections above $\xi =5$. Therefore, discarding these objects does not affect the cosmological results.

\begin{table*}
\centering
\footnotesize
\caption{Cosmological and scaling relation parameters, following the definitions in \cite{Bocquet:2018ukq} and \cite{2016A&A...594A..24P}. We report a brief description and the prior we adopt in our analysis: a range indicates a top-hat prior, while $\mathcal{N}(\mu,\sigma)$ stands for a Gaussian prior with mean $\mu$ and variance $\sigma^2$.}
\label{tab:parameters_ALL}
\begin{tabular}{lcr}
\hline 
Parameter &	Description	& Prior	\\
\hline \vspace{-3mm} \\
\multicolumn{3}{l}{Cosmology}\\
$\Omega_m$ & Matter density & $[0.15,0.4]$ \vspace{0.5mm} \\
$A_s$ & \multirow{2}{*}{\parbox[t]{4cm}{Amplitude of primordial curvature perturbations}} & $[10^{-10},10^{-8}]$ \vspace{0.5mm} \\
    & & \vspace{0.5mm}  \\
$h$ & Expansion rate & $[0.55,0.9]$    \vspace{0.5mm}  \\
$\Omega_b h^2$ & Baryon density & $[0.020,0.024]$ \vspace{0.5mm}  \\
$\Omega_\nu h^2$ &  Massive neutrinos energy density & $[0.0006,0.01]$   \vspace{0.5mm} \\
$n_s$ & Spectral index & $[0.94,1.0]$ \vspace{0.5mm}  \\
\hline \vspace{-3mm}\\

\multicolumn{3}{l}{SPT: SZ scaling relation}\\
$A_\sz$	& Amplitude & $[1,10]$ 	\vspace{0.5mm} \\
$B_\sz$	& Power-law index mass dependence & $[1.2,2]$ \vspace{0.5mm} \\
$C_\sz$	& Power-law index redshift evolution & $ [-1,2] $\vspace{0.5mm} \\
$\sigma_{\ln{\zeta}}$	& Intrinsic scatter	& $[0.01,0.5]$	\vspace{0.5mm} \\
\hline \vspace{-3mm}\\

\multicolumn{3}{l}{SPT: X-ray $Y_X$ scaling relation}\\
$A_{Y_X}$ & Amplitude & $[3,10]$ 	\vspace{0.5mm} \\
$B_{Y_X}$ & Power-law index mass dependence & $[0.3,0.9]$ 	\vspace{0.5mm} \\
$C_{Y_X}$ & Power-law index redshift evolution & $ [-1,5] $\vspace{0.5mm} \\
$\sigma_{\ln{Y_X}}$ & Intrinsic scatter & $[0.01,0.5]$\vspace{0.5mm} \\    
$\de\ln Y_X/\de\ln r$ & Radial slope $Y_X$ profile & $\mathcal{N}(1.12,0.23)$	\vspace{0.5mm} \\    
\hline \vspace{-3mm}\\

\multicolumn{3}{l}{SPT: $\mwl$ scaling relation}\\
$\delta_{\rm{WL,bias}}$ & Coeff. for WL bias & $\mathcal{N}(0,1)$ \vspace{0.5mm} \\
$\delta_{\rm Megacam}$ & Coeff. for error on WL bias  & $\mathcal{N}(0,1)$ \vspace{0.5mm} \\
$\delta_{\rm HST}$ & Coeff. for error on WL bias & $\mathcal{N}(0,1)$ \vspace{0.5mm} \\
$\delta_{\rm{WL,scatter}}$ & Coeff. for lognormal scatter & $\mathcal{N}(0,1)$ \vspace{0.5mm} \\
$\delta_{\rm{WL,LSS}_{\rm Megacam}}$ & Coeff. for normal scatter & $\mathcal{N}(0,1)$ \vspace{0.5mm} \\
$\delta _{\rm{WL,LSS}_{\rm HST}}$ & Coeff. for normal scatter & $\mathcal{N}(0,1)$ \vspace{0.5mm} \\
\hline \vspace{-3mm}\\   

\multicolumn{3}{l}{SPT: Correlation coefficients between scatters}\\
$ \rho(\ln \zeta;\ln M_{\rm{WL}})$ & Correlation coefficient SZ-WL & $[-1,1]$ \vspace{0.5mm} \\
$ \rho(\ln \zeta;\ln Y_X)$ & Correlation coefficient SZ-X & $[-1,1]$ 	\vspace{0.5mm} \\
$ \rho(\ln Y_X;\ln M_{\rm{WL}})$ & Correlation coefficient X-WL & $[-1,1]$ 	\vspace{0.5mm} \\
\hline \vspace{-3mm}\\

\multicolumn{3}{l}{Planck: SZ scaling relation}\\
$\alpha_\sz$ & Power-law index mass dependence & $[1,2.5]$ 	\vspace{0.5mm} \\
$\beta_\sz$	& Power-law index redshift dependence  & Fixed$(0.66)$ or $[0,2]$ 	\vspace{0.5mm} \\
$\sigma_{\log{Y}_\sz}$ & Intrinsic scatter & $\mathcal{N}(0.075,0.01)$ \vspace{0.5mm} \\
$\log{Y_*}_\sz$	& Amplitude	& $\mathcal{N}(-0.186,0.021)$ \vspace{0.5mm} 	\\
$(1-b)_\sz$ & Mass bias & $[0.3,1.3]$\\
\hline \vspace{-3mm}\\
\end{tabular}
\end{table*}

\subsection{Planck satellite}\label{sec:Planck_exp}

The Planck satellite is a mission from the European Space Agency (ESA), which concluded the observations in 2013 \citep{2020A&A...641A...1P}. 
The Planck cluster catalog \citep{2016A&A...594A..27P} is based on full-sky
observations from the 6 channels of the High Frequency Instrument (HFI, \cite{2020A&A...641A...3P}), in the frequency range 100-857 GHz. Similarly to SPT, Planck clusters are extracted using a multi-frequency matched filter technique.  
For the spatial profile of the signal, 
the so-called ``universal pressure profile" from \cite{2010A&A...517A..92A} has been adopted.

The cosmological sample, labeled as ``PSZ2 cosmo", consists of 439 clusters, 433 of which have confirmed redshifts, detected with a signal-to-noise ratio $q>6$, on the $65\%$ of the sky remaining after masking high dust emission regions and point sources.
The signal-to-noise ratio is defined as 
\begin{equation}\label{eq:q_gen}
q = \dfrac{Y_{500}}{\sigma_{\text{f}}(\theta_{500},l,b)} \, ,
\end{equation}
where $Y_{500}$ is the integrated compton parameter (tSZ signal for a cluster) and $\sigma_{\text{f}}(\theta_{500},l,b)$ is the detection filter noise as a function of the cluster angular size, $\theta_{500}$, and sky position in galactic coordinates $(l,b)$.
The PSZ2 cosmo sample spans the mass range $M_{\text{SZ}} = (2-10)\times 10^{14} M_{\odot}$ and the redshift range $z = [0, 1]$. 

The Planck cosmological analysis is based on a Poissonian likelihood, constructed on counts of redshift and signal-to-noise ratio:
\begin{equation}\label{eq:planck_like1}
\ln{\mathscr{L}_\text{P}} = \sum _{i,j}^{N_zN_q}\left[ N_{ij}\ln{\bar{N}_{ij}} - \bar{N}_{ij} - \ln{(N_{ij}!)} \right] \, .    
\end{equation}
In the above equation, $N_z$ and $N_q$ are the total number of redshift and signal-to-noise bins, with redshift binning $\Delta z = 0.1$ and signal-to-noise ratio binning $\Delta \log q = 0.25$.
$N_{ij}$ represents the observed number counts of cluster. $\bar{N}_{ij}$ is the predicted mean number of objects in each bin, modelled by theory as
\begin{equation}\label{eq:planck_like2}
\bar{N}_{ij} = \dfrac{dN}{dzdq}(z_i,q_j)\Delta z \Delta q \, .
\end{equation}

We report here the main steps to evaluate the theoretical cluster number counts and refer to \cite{2016A&A...594A..24P} for the complete description. 
The cluster distribution can be written as
\begin{equation}\label{eq:planck_NC1}
    \dfrac{dN}{dzdq} = \iint d\Omega \, dM_{500} \, \dfrac{dN}{dzdM_{500}d \Omega} P[q|\bar{q}_m(M_{500},z,l,b)] \, ,
\end{equation}
where
\begin{equation}\label{eq:planck_NC2}
\dfrac{dN}{dzdM_{500}d \Omega} = \dfrac{dV}{dzd\Omega} \dfrac{dN}{dVdM_{500}}
\end{equation}
is the product of the volume element and the halo mass function respectively.

In Eq.~\ref{eq:planck_NC1}, the quantity $P[q|\bar{q}_m(M_{500},z,l,b)]$ represents the distribution of the signal-to-noise ratio $q$ given the mean value $\bar{q}_m(M_{500},z,l,b)$, predicted by the model, for a cluster located at position $(l,b)$, with mass $M_{500}$ and redshift $z$.
The $P[q|\bar{q}_m]$ distribution takes into account the noise fluctuations and the intrinsic scatter $\sigma_{\ln{Y}}$ 
of the actual cluster signal
$Y_{500}$ around the mean value, $\bar{Y}_{500}(M_{500}, z)$, predicted from the scaling relation. In this analysis, we assume that the intrinsic scatter does not show any dependence on $(M_{500},z)$, following the original approach in \cite{2016A&A...594A..24P}.

The relation between the cluster observables $Y_{500}, \, \theta_{500}$ and the cluster mass and redshift is described by a log-normal distribution function $P(\ln{Y_{500}},\theta_{500}|M_{500},z)$. 
The mean values of this distribution are given by the scaling relations $\bar{Y}_{500}(M_{500},z)$ and $\bar{\theta}_{500}(M_{500},z)$, defined as

\begin{eqnarray}\label{eq:planck_SR1}
    E^{-\beta_{\text{SZ}}}(z)\left[ \dfrac{D_A^2(z) \bar{Y}_{500}}{10^{-4} \text{Mpc}^2} \right] &=& Y_{*,\text{SZ}} \left[ \dfrac{h}{0.7} \right]^{-2+\alpha_{\text{SZ}}} \notag \\
    &\times& \left[ \dfrac{(1-b)_{\text{SZ}}M_{500}}{6 \times 10^{14}M_{\odot}} \right]^{\alpha_{\text{SZ}}}
\end{eqnarray}

\begin{eqnarray}\label{eq:planck_SR2}
    \bar{\theta}_{500} &=& \theta_* \left[ \dfrac{h}{0.7} \right]^{-2/3} \left[ \dfrac{(1-b)_{\text{SZ}}M_{500}}{6 \times 10^{14}M_{\odot}} \right]^{1/3} \notag \\
     &\times& E^{-2/3}(z) \left[ \dfrac{D_A(z)}{500 \text{Mpc}} \right]^{-1} \, .
\end{eqnarray}
In the above equations, $D_A(z)$ is the angular diameter distance and $E(z) \equiv H(z)/H_0$. 

In the original analysis of \cite{2014A&A...571A..20P,2016A&A...594A..27P}, 
the calibration of Eqs.~\ref{eq:planck_SR1} and \ref{eq:planck_SR2} is based on X-ray observations of 71 clusters under the assumption of hydrostatic equilibrium. To account for possible deviations from this assumption (due to cluster physics, observational effects or selection effects), the mass bias parameter $b$ is introduced in the analysis, such that the relation between the HE mass ($M_{\text{SZ}}$) and the real cluster mass is $M_{\text{SZ}} = (1-b) M_{500}$.

In order to evaluate the mass bias (and therefore the real cluster mass), WL mass determinations are introduced in the analysis. For the baseline cosmological analysis, Planck collaboration adopts the evaluation from the Canadian Cluster Comparison Project (\cite{Hoekstra:2015gda}, CCCP hereafter), $(1-b)_{\text{SZ}}=0.780 \pm 0.092$, based on 20 clusters.
We stress that the mass bias is considered as a constant quantity, i.e. not allowing for dependence on the cluster mass and redshift. 
The original values for the scaling relation parameters (from X-ray and WL calibration) are reported in Table~\ref{tab:planck_SR1}, following \cite{2016A&A...594A..24P}. 
We note that as baseline we assume the self-similarity model for 
the redshift evolution of the cluster population.
This translates into fixing the $\beta$ parameter to $\beta_{\text{SZ}} = 2/3$.

\begin{deluxetable}{cc|cc}
\tablecaption{Original calibration of Planck scaling relation parameters. $\mathcal{N}(\mu,\sigma)$ stands for a Gaussian prior with mean $\mu$ and variance $\sigma^2$.}
\label{tab:planck_SR1}
\tablewidth{10pt}
\tablehead{
\colhead{Parameter} & \colhead{Value} & \colhead{Parameter} & \colhead{Value}}
\startdata
$\log{Y_{*,\text{SZ}}}$ & $\mathcal{N}(-0.19,0.02)$ & $\sigma_{\ln{Y_{\text{SZ}}}} ^a$ & $\mathcal{N}(0.173,0.023)$ \\
$\alpha_{\text{SZ}}$ & $\mathcal{N}(1.79,0.08)$ & $(1-b)_{\text{SZ}}$ & $\mathcal{N}(0.780,0.092)$ \\
$\beta_{\text{SZ}}$ & $0.66$ &  &\\
\enddata
\tablecomments{$^{a}$ In practice, in the analysis we use the parameter $\sigma_{\log{Y_{\text{SZ}}}} = \mathcal{N}(0.075,0.01)$.}
\end{deluxetable}

In summary, the main difference between Planck and SPT mass calibrations lies in the 
use of external data (from other cluster samples) for Planck vs. the use of internal data (direct follow-up observations) for SPT.
Therefore, when analyzing Planck data, it is possible to relax some of the external calibration results and provide independent constraints on some of the scaling relation parameters.

\section{Method} \label{sec:method}

In this section we describe the strategy that we adopted to combine Planck and SPT data, in order to avoid covariance between the two samples. In particular, we discuss how we modify the original Planck likelihood to provide a proper combination with the SPT one.
Finally, we describe the method we use to provide a new evaluation of Planck cluster masses.

\subsection{Combining Planck and SPT cluster likelihoods}\label{sec:comb_like}

In order to combine Planck and SPT cluster likelihoods, it is necessary to take into account the overlapping area of the observed sky and the clusters in common between the two catalogs.
We choose to modify the Planck likelihood. In particular, we perform a split in redshift of the entire likelihood. For $z \leq 0.25$, where we do not have cluster data from the SPT-SZ survey, we rely on the original version for the Planck likelihood. For $z >0.25$, we modify the Planck likelihood, removing the part of the sky observed also by the SPT-SZ survey, and the clusters in common with the SPT-SZ catalog. Hereafter, we refer to this new Planck redshift-splitted likelihood as ``PvSPLIT". With this choice, we can therefore treat the two Planck and SPT cluster likelihoods independently.

We now discuss in details the approach used to build the $z>0.25$ part of the likelihood. 
From the original Planck analysis \citep{2016A&A...594A..27P}, the cosmological cluster catalog is built through the application of a multi-frequency matched filter technique to the HFI frequency maps, selecting objects with $S/N>6$. The detection algorithm first divides the sky in 504 tangential patches of $10^{\circ} \times 10^{\circ} $ area, with constant values of detection noise. After applying the galactic and point source mask, we are left with 417 sky patches, covering $\sim 65\%$ of the sky. Cluster candidates are then detected in each sky patch: the final catalog is therefore completely dependent on the characteristics of the detection process, including the sky patches division. When modifying the Planck cluster likelihood for $z>0.25$, we therefore need to keep this patches configuration.

\begin{figure}
  \centering
  \includegraphics[scale=0.4]{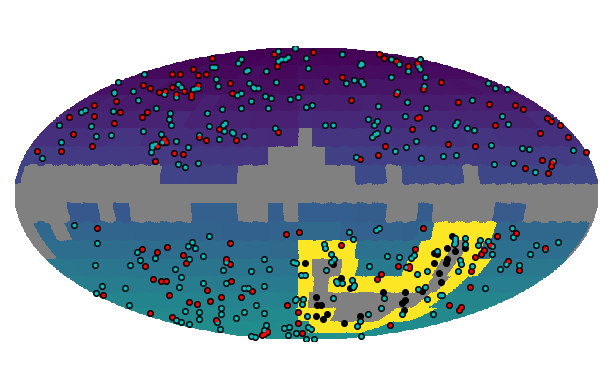} \\
 \includegraphics[scale=0.5]{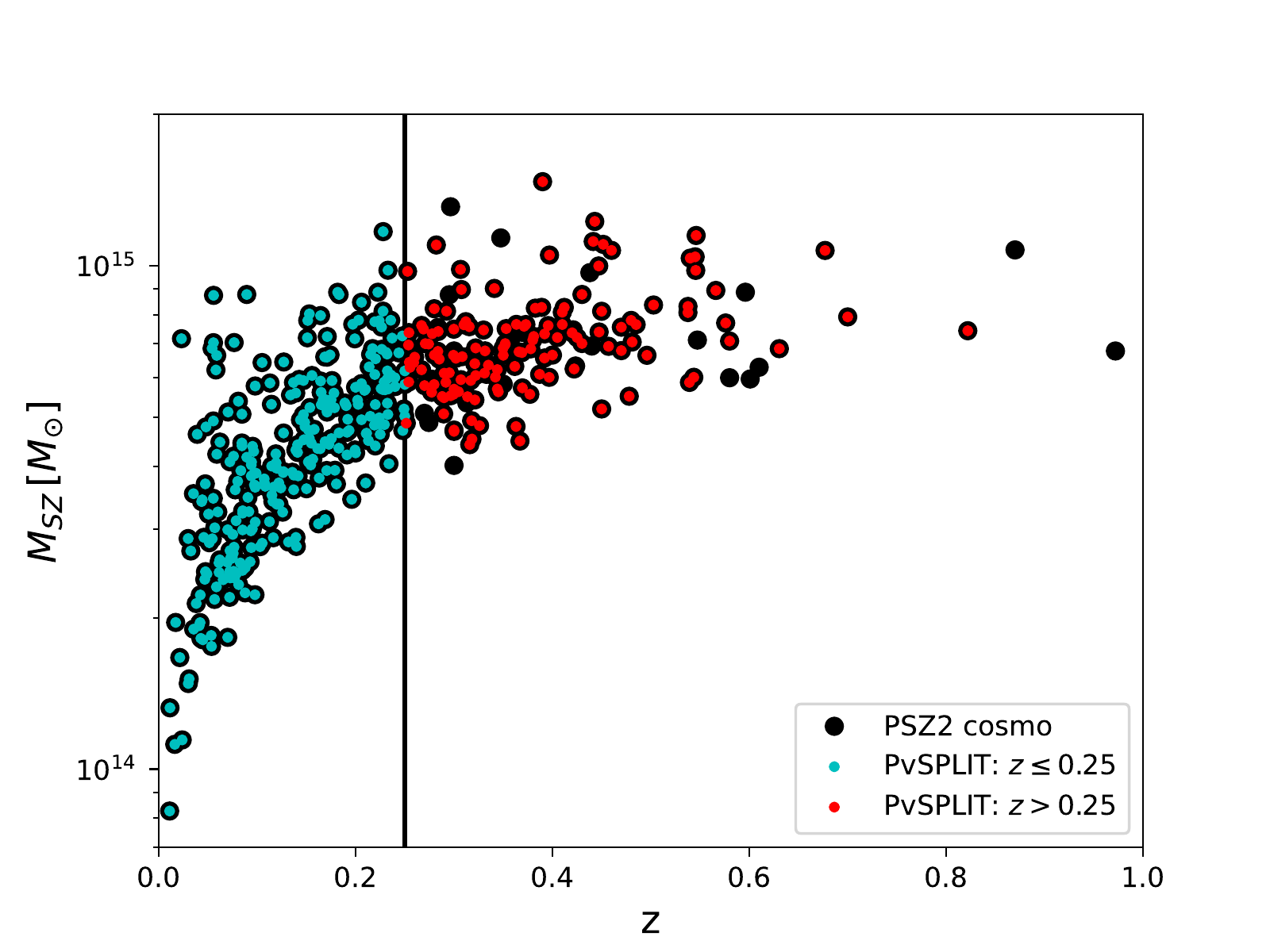}
  \caption{\textit{Upper panel.} Map of Planck patches in galactic coordinates. In grey we show the removed patches, due to the Planck galactic mask and the fully overlapped area with SPT observations. In yellow we highlight the 35 partly overlapping patches between Planck and SPT. \textit{Upper and lower panel.} Comparison between the PSZ2 cosmo catalog (black points), with the PvSPLIT catalog. In the upper panel, we show how the clusters are distributed in the sky. In the lower panel, we show the mass-redshift distribution, considering $M_{\text{SZ}}$ mass from \cite{2016A&A...594A..24P}. The cyan points are the clusters considered in the $z \leq 0.25$ part of the PvSPLIT likelihood. The red points are the clusters considered in the $z>0.25$ PvSPLIT likelihood.}
  \label{fig:planck_vsplit1}
\end{figure}

We identify 16 patches that fully overlap with the SPT observed sky. We remove those patches from the sky area in the likelihood. Furthermore, we identify 35 patches with a partial overlapping between Planck and SPT sky. In this case, we decide to keep them in the analysis, but reduce the sky fraction in each patch, according to the area that is actually observed by both experiments. The remaining observed sky is shown in Fig.~\ref{fig:planck_vsplit1}, upper panel. We show in grey the removed patches, due to Planck galactic mask and the Planck-SPT fully overlapped area. In yellow, we highlight the patches that partly overlaps between Planck and SPT-SZ survey. 

For the cluster catalog, 
we remove 27 clusters in common with the 
SPT-SZ cosmological catalog and 2 clusters that fall in the removed patches.
We also introduce redshifts for the 6 clusters whose redshifts was unknown in the original PSZ2 cosmo sample. 
We report the new redshifts in Table~\ref{tab:planck_z}, specifying if these values have been obtained from photometric (P) or spectroscopic (S) observations. We show the new cluster distribution in Fig.~\ref{fig:planck_vsplit1}: in the upper panel we show how Planck clusters are distributed on the observed sky, in the lower panel we show the mass-redshift cluster distribution. 

\begin{deluxetable*}{ccc}
\tablecaption{Redshifts for clusters without redshifts in the original PSZ2 cosmological catalog obtained from photometric (P) or spectroscopic (S) observations.\label{tab:planck_z}}
\tablewidth{10pt}
\tablehead{
\colhead{Cluster} & \colhead{$z$} & \colhead{Reference}}
\startdata
PSZ2 G011.36-72.93 & $z = 0.63 \pm 0.04$ (P) & \cite{Bleem:2019fvp} \\
PSZ2 G107.83-45.45 & $z = 0.55 \pm 0.05$ (P) & \cite{Boada:2018ekk} \\
PSZ2 G160.83-70.63 & $z = 0.24 \pm 0.03$ (P) & \cite{Aguado-Barahona:2019wvv}\\
PSZ2 G237.41-21.34 & $z = 0.31 \pm 0.04$ (P) & this work $^a$\\
PSZ2 G293.01-65.78 & $z = 0.206 \pm 0.006 $ (P) & \cite{2019MNRAS.488..739K} \\
PSZ2 G329.48-22.67 & $z = 0.249 \pm 0.001$ (S) & \cite{Amodeo:2017xeh}\\
\enddata
\tablecomments{$^{a}$ from Pan-STARRS \citep{2016arXiv161205560C} following \cite{Bleem:2019fvp}}
\end{deluxetable*}

Following Eq.~\ref{eq:planck_like1}, the new Planck PvSPLIT likelihood therefore reads
\begin{eqnarray}\label{eq:pvsplit_like1}
\ln{\mathscr{L}_\text{P}} &=& \ln{\mathscr{L}_{\text{P}1}} + \ln{\mathscr{L}_{\text{P}2}} \notag \\
 &=& \sum _{i_1,j}^{N_{z_1} N_q}\left[ N_{i_1 j}\ln{\bar{N}_{i_1 j}} - \bar{N}_{i_1 j} - \ln{(N_{i_1 j}!)} \right] \notag \\
 &+& \sum _{i_2,j}^{N_{z_2} N_q}\left[ N_{i_2j}\ln{\bar{N}_{i_2j}} - \bar{N}_{i_2j} - \ln{(N_{i_2j}!)} \right]
\end{eqnarray}

\noindent where we adopt a binning in redshift of $\Delta z = 0.05$, such that we have $N_{z_1} = 5$ redshift bins up to $z\leq 0.25$, and $N_{z_2} = 15$ above. For the binning in the signal-to-noise ratio, we follow the original analysis, with $\Delta \log{q} = 0.25$.  
The total likelihood for the combined analysis of Planck and SPT, following Eqs.~\ref{eq:spt_likelihood1} and \ref{eq:pvsplit_like1}, is therefore defined as
\begin{equation}
\ln{\mathscr{L}_{\text{TOT}}} = \ln{\mathscr{L}_{\text{SPT}}} + \ln{\mathscr{L}_{\text{P}1}} + \ln{\mathscr{L}_{\text{P}2}} \, .
\end{equation}

\subsection{Sampling recipe}\label{sec:mcmc}

For the cosmological analysis, 
we make use of the complete SPT likelihood, described in section \ref{sec:SPT}. In particular, we rely on the combination of the SPT-selected clusters with their detection significance and redshift, together with the WL and X-ray follow-up data, where available. 
Following the definition in \cite{Bocquet:2018ukq}, we refer to this data set as ``SPTcl" (SPT-SZ + WL + $Y_X$).

For the Planck part of the likelihood, we use the PvSPLIT version described in the previous section. We adopt the parametrisation for the scaling relations described in Eqs.~\ref{eq:planck_SR1} and \ref{eq:planck_SR2}. 

In this analysis, we want to test the capability of the Planck+SPT combination to constrain the Planck scaling relation parameters. For this reason, we do not consider the original X-ray+WL calibration reported in table~\ref{tab:planck_SR1} when analyzing Planck data.
As a baseline, we use the X-ray calibration for the $\log{Y_{*,\sz}}$ and $\sigma_{\log{Y}_\sz}$ parameters, as reported in Table~\ref{tab:planck_SR1}, and we assume the self-similarity model for the cluster evolution, i.e. $\beta_\sz = 0.66$. We therefore focus the analysis on the constraints that we can obtain on the mass bias and the power-law index of the mass dependence, $(1-b)_\sz$ and $\alpha_\sz$. We refer to this parameter exploration and likelihood combination as the baseline ``SPTcl + PvSPLIT" results.
As a further test, we also relax the assumption of redshift self-similar evolution and let the $\beta_\sz$ parameter free to vary.

For the cosmological parameters, we assume a $\nu \Lambda$CDM scenario. We vary the following parameters: the total matter density $\Omega_m$, the amplitude of primordial curvature perturbation $A_s$, the Hubble rate $h$, the baryon density $\Omega_bh^2$, the spectral index for scalar perturbations $n_s$ and the massive neutrino energy density $\Omega_\nu h^2$.
When providing the results for the cosmological parameters, we focus also on the $\sigma_8$ quantity.
We report all the parameters, with the priors used in the analysis, in Table \ref{tab:parameters_ALL}.
The sampling of the likelihood is performed with the importance nested sampler algorithm {\tt MultiNest} \citep{Feroz:2008xx}, within the {\tt cosmoSIS} package \citep{Zuntz:2014csq}. 

As shown in section \ref{sec:data}, the halo mass function is a fundamental ingredient for the evaluation of the cluster number counts. For both the SPT and Planck parts of the analysis, we make use of the evaluation from \cite{Tinker:2008ff}.

\subsection{Mass evaluation}\label{sec:me}

We now describe the approach we use to provide a new evaluation of true Planck cluster masses, $M_{500}$. 
We follow the discussion in \cite{2016A&A...594A..24P}.
We start from the Planck cluster observable, the signal-to-noise ratio $q$, and evaluate $P(M_{500}|q)$, which represents the conditional probability that a cluster with given signal-to-noise ratio $q$ has a mass $M_{500}$.
Following the Bayes theorem, this probability is defined as
\begin{equation}\label{eq:bayes_m}
P(M_{500}|q) \propto P(q|M_{500}) \, P(M_{500})
\end{equation}
where the first term is the conditional probability of the data (the signal-to-noise ratio $q$), given the model (the cluster mass $M_{500}$), and the second term is the mass probability distribution.
The latter is related to the mass function $dN/dM_{500}$, such that
\begin{equation}
P(M'_{500}) = \dfrac{dN/dM_{500}|_{M'_{500}}}{\int dM_{500} \, dN/dM_{500}} \, .
\end{equation}

In order to evaluate $P(q|M_{500})$, we follow the recipe for $P[q|\bar{q}_m(M_{500},z,l,b)]$, that represents the probability distribution of the observed signal-to-noise ratio $q$ given the mean one, $\bar{q}_m$, as already mentioned in section \ref{sec:data}. Following Eq.~\ref{eq:q_gen}, the mean theoretical signal-to-noise ratio is defined as
\begin{equation}\label{eq:q_mm}
\bar{q}_m \equiv \dfrac{\bar{Y}_{500}}{\sigma_{\text{f}}(\bar{\theta}_{500},l,b)} \, ,
\end{equation}
where $\bar{Y}_{500}$ and $\bar{\theta}_{500}$ are the mean values of the scaling relations defined in Eqs.~\ref{eq:planck_SR1} and \ref{eq:planck_SR2} and $\sigma_{\text{f}}$ is the detection filter noise. For fixed values of the cosmological and scaling relation parameters, we have therefore a unique relation between the cluster mass $M_{500}$ and $\bar{q}_m$.

The probability distribution can be evaluated as
\begin{eqnarray}\label{eq:Pq_qmm}
P[q|\bar{q}_m(M_{500},z,l,b)] &=& \int d \ln q_m \dfrac{e^{-(q-q_m)^2/2}}{\sqrt{2\pi}} \notag \\
 &\times & \dfrac{e^{-\ln^2{(q_m/\bar{q}_m})/2\sigma^2_{\ln Y}}}{\sqrt{2\pi}\sigma_{\ln Y}} \, .
\end{eqnarray}
In the above equation, the second term accounts for the intrinsic scatter of the mass-observable relations, while the first term links the theoretical signal-to-noise ratio $q_m$ to the observed one, assuming pure Gaussian noise.

In practice, we adopt a Monte-Carlo extraction based approach, starting from the parameter space exploration performed for the SPTcl + PvSPLIT analysis.
For a given cosmological and scaling relation model, we extract $M_{500}$ in the range $[3\cdot 10^{13},1.2\cdot10^{16}] [M_{\odot}h^{-1}]$ (following what is done in the PvSPLIT likelihood)
according to the halo mass function distribution. We then evaluate $\bar{Y}_{500}$, $\bar{\theta}_{500}$ and consequently $\bar{q}_m$, following Eq.~\ref{eq:q_mm}.

For the given mean theoretical signal-to-noise ratio, we then extract $q_m$, following a log-normal distribution with standard deviation equal to the intrinsic mass-observable relation scatter, $\sigma_{\ln Y}$.

Given $q_m$, we can extract the estimate of the observed signal-to-noise ratio $q_{\text{est}}$, following a Gaussian distribution with standard deviation equal to 1.
We then select $N$ values of $q_{\text{est}}$ around the corresponding observed signal-to-noise $q$, therefore selecting the corresponding values of $M_{500}$.
The posterior distributions for $M_{500}$ are obtained marginalizing over the full parameter space, considering cosmological and scaling relation parameters. The resulting catalog therefore provides the first sample of Eddington-bias-corrected calibrated cluster masses,
that include correlations associated with scaling relation and cosmological parameters. As detailed in section \ref{sec:mass_massbias_results}, we make this catalog publicly available.

\subsubsection{Mass bias}\label{sec:mass_bias}

With the evaluation of $M_{500}$, 
we can estimate directly and 
for each cluster in the PSZ2 cosmo sample, the mass bias 
as $(1-b)_M = M_{\text{SZ}}/M_{500}$. 
We use $M_{\text{SZ}}$ estimations provided by the Planck collaboration \citep{2016A&A...594A..24P} for the 433 clusters in PSZ2 cosmo for which the redshift was originally provided. 
In practice, we expand the procedure for $M_{500}$ evaluation presented in previous section. For each cluster, at each step of the Monte-Carlo extraction, we also extract $M_{\text{SZ}}$ within the constraints of the Planck measurements. We then evaluate $(1-b)_M$. The final constraints on $(1-b)_M$ are therefore obtained marginalizing over cosmological and scaling relation parameters, and take into account the uncertainty on $M_{\text{SZ}}$.

We highlight the difference between the scaling relation parameter $(1-b)_{\text{SZ}}$ entering Eqs.~\ref{eq:planck_SR1} and \ref{eq:planck_SR2}, and the quantity we investigate here. 
The assumptions of spherical collapse, hydrostatic equilibrium and self-similarity lead to the formulation of Eqs.~\ref{eq:planck_SR1}, \ref{eq:planck_SR2} that link the tSZ observables and the cluster mass. In this case, the mass bias $(1-b)_{\text{SZ}}$ is introduced to take into account any generic departure from hydrostatic equilibrium. 
Nevertheless, as discussed in \cite{2016A&A...594A..27P}, 
$M_{\text{SZ}}$ is evaluated as the real cluster mass
combining the scaling relation information with the output of the matched filter approach used to detect the clusters. The combination of these different approaches might select cluster scales that do not actually maximise the S/N ratio from the matched filter algorithm, and therefore introduce a further bias in the estimation of the real cluster mass. 

We attempt therefore to provide a complete characterization of the ``measured" mass bias $(1-b)_M$, analyzing the dependencies with respect to theoretical modelling and observational assumptions. 

We consider a mass and redshift evolution for the mass bias. The goal 
is to understand if we need to further improve the theoretical modelling of the scaling relations. Indeed, as discussed e.g. in \cite{Salvati:2019zdp} comparing recent WL mass calibrations \citep{vonderLinden:2014haa,Okabe:2015faa,Sereno:2015jda,Smith:2015qhs,Penna-Lima:2016tvo,Sereno:2017zcn,Herbonnet:2019byy}, 
a correct calibration of the mass bias might need to take into account the mass and redshift distribution of the full cluster catalog. 

In addition, we analyze a possible link between the evaluation of the mass bias (and therefore of the cluster mass) and the cluster position in the sky. This dependence might be related to the observational strategy, as well as to the 
assumptions for the ingredients used in the matched filter approach. 
As discussed in section \ref{sec:comb_like}, the Planck sky area, used for the cluster cosmological analysis, is divided into 417 patches, with each patch having a different value of detection noise $\sigma_{\text{f}}(\theta_{500},l,b)$. This noise depends on the filter size $\theta_{500}$ and is therefore related to the matched filter approach used to detect clusters in the Planck map. 
Therefore, the analysis of a possible dependence of the mass bias with respect to the detection noise allows us to quantify the systematic uncertainties coming from the modelling of the whole selection approach. 

Considering the mass, redshift and noise dependence, we define the theoretical mass bias $(1-b)_M^{\text{th}}$ as
\begin{eqnarray}\label{eq:bias_Mzn}
   (1-b)_M^{\text{th}} &=& A_{\text{bias}} \, \left( \dfrac{M_{500}}{M_*}\right)^{\gamma_M} \, \left( \dfrac{1+z}{1+z_*} \right)^{\gamma_z} \notag \\
   &\times &  \left( \dfrac{\sigma_{\text{f}}(\theta_{500},l,b)}{\sigma_{\text{f},*}(\theta_{500})}\right)^{\gamma_n}
\end{eqnarray}

\noindent where $M_* = 4.68 \cdot 10^{14} \, M_{\odot} h^{-1} $ is the median mass of the sample (obtained from our analysis), $z_* = 0.21$ is the median redshift of the sample and $ \sigma_{\text{f},*}(\theta_{500})$ is the median detection noise at the given $\theta_{500}$.
In order to get constraints on the mass, redshift and detection noise dependence, we 
perform a fit between the measured Planck $M_{\text{SZ}}$ masses and a theoretical estimation $M_{\text{SZ}}^{\text{th}}$ defined as 
\begin{equation}\label{eq:msz_th}
M_{\text{SZ}}^{\text{th}} =(1-b)_M^{\text{th}} \cdot M_{500} \, ,
\end{equation}
where the masses $M_{500}$ are derived following the method described in section~\ref{sec:me}.

\section{Results} \label{sec:results}
In this section we report the results for the combined cosmological analysis of Planck and SPT cluster likelihoods. We also provide an estimate of the cluster mass and mass bias for Planck PSZ2 cosmological sample.

\subsection{Cosmological and scaling relation parameters}\label{sec:results_cosmo}
The results presented in this analysis are obtained combining the full SPT likelihood with the new Planck likelihood, presented in section \ref{sec:data} and \ref{sec:method}, SPTcl + PvSPLIT. When discussing our results, we focus on the constraints for the cosmological parameters $\Omega_m$ and $\sigma_8$ and for the Planck scaling relation parameters $(1-b)_{\text{SZ}}$ and $\alpha_{\text{SZ}}$. We start by comparing the results for SPTcl + PvSPLIT baseline combination with constraints obtained when considering the SPT data and Planck data alone.

We stress that, when providing results for Planck data alone, we are actually considering the combination of cluster counts with measurements of BAO
\citep{Alam:2016hwk,Beutler:2012px,10.1111/j.1365-2966.2011.19250.x,Ross:2014qpa}, together with constraints on the baryon density $\Omega_b h^2$ from Big Bang Nucleosynthesis (BBN hereafter).
We also consider the full calibration of the scaling relation parameters (as reported in Table~\ref{tab:planck_SR1}), following the analysis in \cite{2016A&A...594A..24P}. 
In this work, we simply perform a new analysis using the {\tt MultiNest} sampler, within the {\tt cosmoSIS} package, in order to provide consistent results. This dataset combination is labelled as ``PvFULL".

\begin{figure*}
\epsscale{0.85}
\plotone{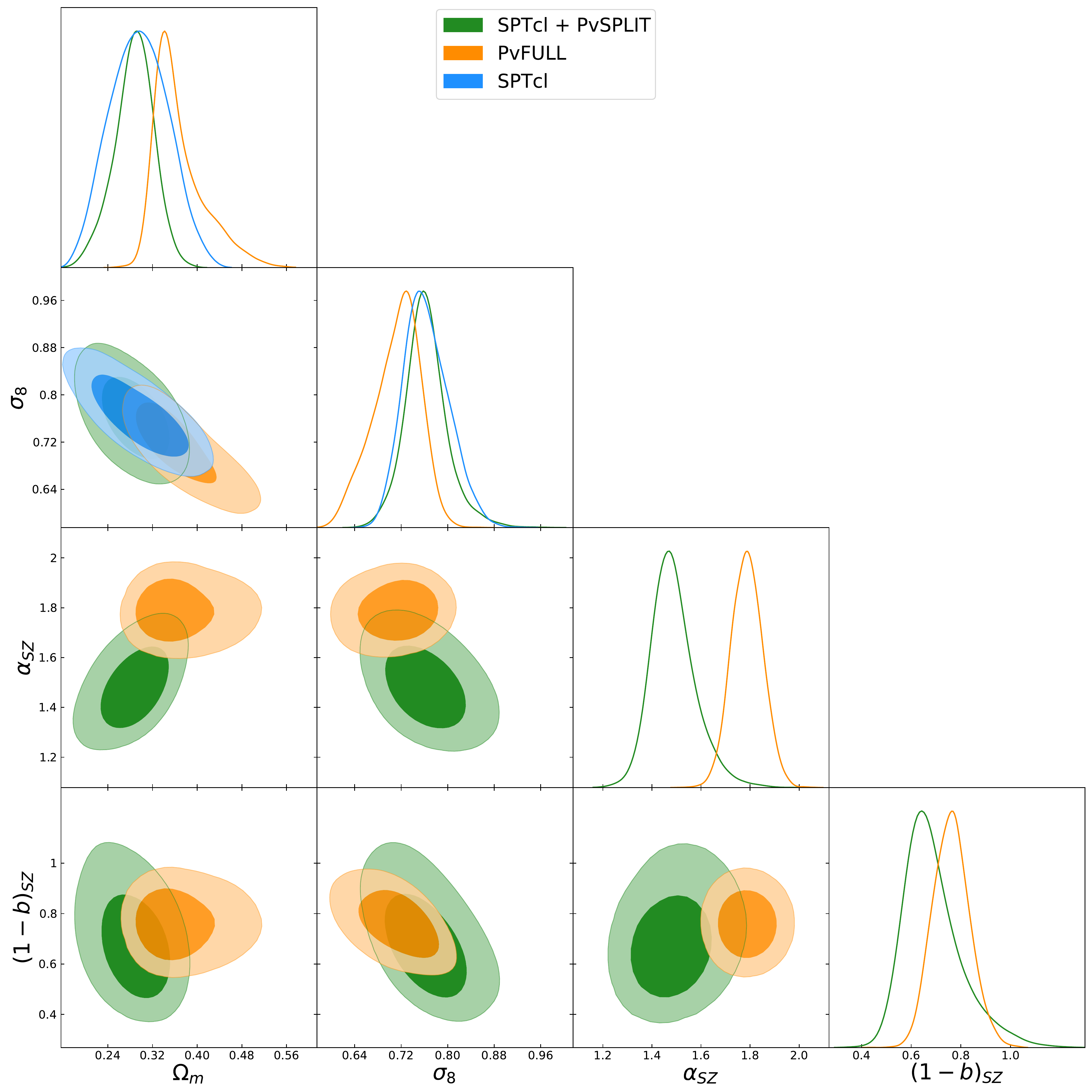}
\caption{We show the one-dimensional and two-dimensional probability distributions for the cosmological ($\Omega_m$, $\sigma_8$) and Planck scaling relation ($\alpha_{\text{SZ}}$, $(1-b)_{\text{SZ}}$) parameters. The contours represent the $68\%$ and $95\%$ CL. We compare results for different dataset combinations: SPTcl + PvSPLIT in green (baseline results of this analysis), PvFULL in orange and SPTcl in blue. We refer to the text for the complete description of the datasets. \label{fig:baseline_tri}}
\end{figure*}

We report the constraints on cosmological and scaling relation parameters in Table~\ref{tab:all_results}. We show the $68\%$ confidence level (CL hereafter) constraints for all the parameters. In the triangular plot in Fig.~\ref{fig:baseline_tri} we show the one-dimensional and two-dimensional probability distributions for the cosmological and scaling relation parameters for the main comparison between the baseline SPTcl + PvSPLIT and the original PvFULL and SPTcl analysis.

\begin{deluxetable*}{ccccccc}
\tablecaption{We report the $68\%$ CL constraints on cosmological and scaling relation parameters for different dataset combinations. We refer to the text for the full dataset description. The MCMC chain the SPTcl + PvSPLIT combination is available at \url{https://pole.uchicago.edu/public/data/sptplanck\_cluster.} \label{tab:all_results}}
\tablewidth{10pt}
\startdata
\tablehead{
\colhead{Parameter} & \multicolumn6c{$\nu \Lambda$CDM}} 
 & \colhead{SPTcl + PvSPLIT} &\colhead{PvFULL} & \colhead{SPTcl}
 & \colhead{PvSPLIT} & \colhead{SPTcl + PvSPLIT + $\ln (10^{10} A_s)$ } &\colhead{SPTcl + PvSPLIT + $\beta_{\text{SZ}}$}
 \\
\vspace{0.15cm}
$\Omega_m$ &  $0.29 _{-0.03}^{+0.04}$ & 
$0.37 _{-0.06}^{+0.02}$ & 
$0.30 \pm 0.03$ &
$ 0.38 _{-0.06}^{+0.02}$ &
$ 0.30 _{-0.04}^{+0.03}$ &
$ 0.28 _{-0.04}^{+0.03}$
\\
\vspace{0.15cm}
$\sigma_8$ & $ 0.76_{-0.04}^{+0.03}$  & 
$ 0.71 _{-0.03}^{+0.05}$ & 
$0.76 _{-0.04}^{+0.03}$ &
$0.68 _{-0.03}^{+0.04}$ &
$ 0.75 \pm 0.03$  &
$ 0.77 \pm 0.04$
\\
\vspace{0.15cm}
$H_0$ & $ 61.3 _{-6.3}^{+1.3}$  & 
$ 71.0 _{-4.0}^{+1.6}$ & 
$61.5 _{-6.0}^{+2.6}$ &
$71.2 _{-4.0}^{+1.7}$ &
$ 69.4 _{-14.4}^{+5.9}$  &
$ 61.8 _{-6.8}^{+1.3}$
\\
\vspace{0.15cm}
$\alpha_{\text{SZ}}$ & $ 1.49 _{-0.10}^{+0.07}$  &  
$1.79 \pm 0.06$ & 
$-$ &
$ 1.71 _{-0.09}^{+0.07}$&
$ 1.60 _{-0.18}^{+0.10}$ &
$ 1.48 _{-0.10}^{+0.07}$ 
\\
\vspace{0.15cm}
$(1-b)_{\text{SZ}}$ & $ 0.69 _{-0.14}^{+0.07}$  & 
$0.76 _{-0.08}^{+0.07}$ & 
$ -$ &
$ 0.79 \pm 0.07$&
$ 0.74_{-0.16}^{+0.09}$ &
$ 0.71 _{-0.14}^{+0.08}$
\\
\vspace{0.15cm}
$\beta_{\text{SZ}}$ & $ 0.67$  & 
$ 0.67$  & 
$ -$ &
$ 0.67$ &
$ 0.67$  &
$ 0.57 _{-0.51}^{+0.20} $
\\
\enddata
\end{deluxetable*}

From these results, we see that SPT cluster data are driving the constraining power, as it is shown from the shift of $\Omega_m$ contours towards lower values and $\sigma_8$ contours towards larger values for the SPTcl + PvSPLIT baseline combination, with respect to PvFULL constraints. 
We stress again that for the SPTcl + PvSPLIT baseline combination we are not including BAO+BBN dataset and part of the X-ray+WL mass calibrations when considering Planck data, therefore losing part of the constraining power that leads to the tight bounds obtained for the PvFULL analysis (as further discussed in Appendix \ref{sec:app1}). 

We now focus on the Planck scaling relation parameters $(1-b)_{\text{SZ}}$ and $\alpha_{\text{SZ}}$.
Regarding the mass bias, we find $(1-b)_{\text{SZ}} = 0.69 _{-0.14}^{+0.07}$. Although pointing towards low value of $(1-b)$, this result is still consistent with constraints obtained from recent WL calibration and numerical simulation analyses, see e.g. a collection of results in \cite{Salvati:2017rsn} and \cite{Gianfagna:2020que}. Nevertheless, not considering the WL calibration from the CCCP analysis (used in the original Planck analysis) leads to a slight enlargement in the constraints. 

Regarding the mass slope $\alpha_{\text{SZ}}$, we find $\alpha_{\text{SZ}} = 1.49 _{-0.10}^{+0.07}$, which is $\sim 4 \, \sigma$ away with respect to the value obtained when adopting the X-ray calibration, $\alpha_{\text{SZ}} = 1.79 \pm 0.06$. 
We recall here that, following the definition of the scaling relations in Eqs.~\ref{eq:planck_SR1} and \ref{eq:planck_SR2}, the value of $\alpha_{\text{SZ}} \simeq 1.8$ is in agreement with self-similarity assumption.

\begin{figure*}
\epsscale{0.85}
\plotone{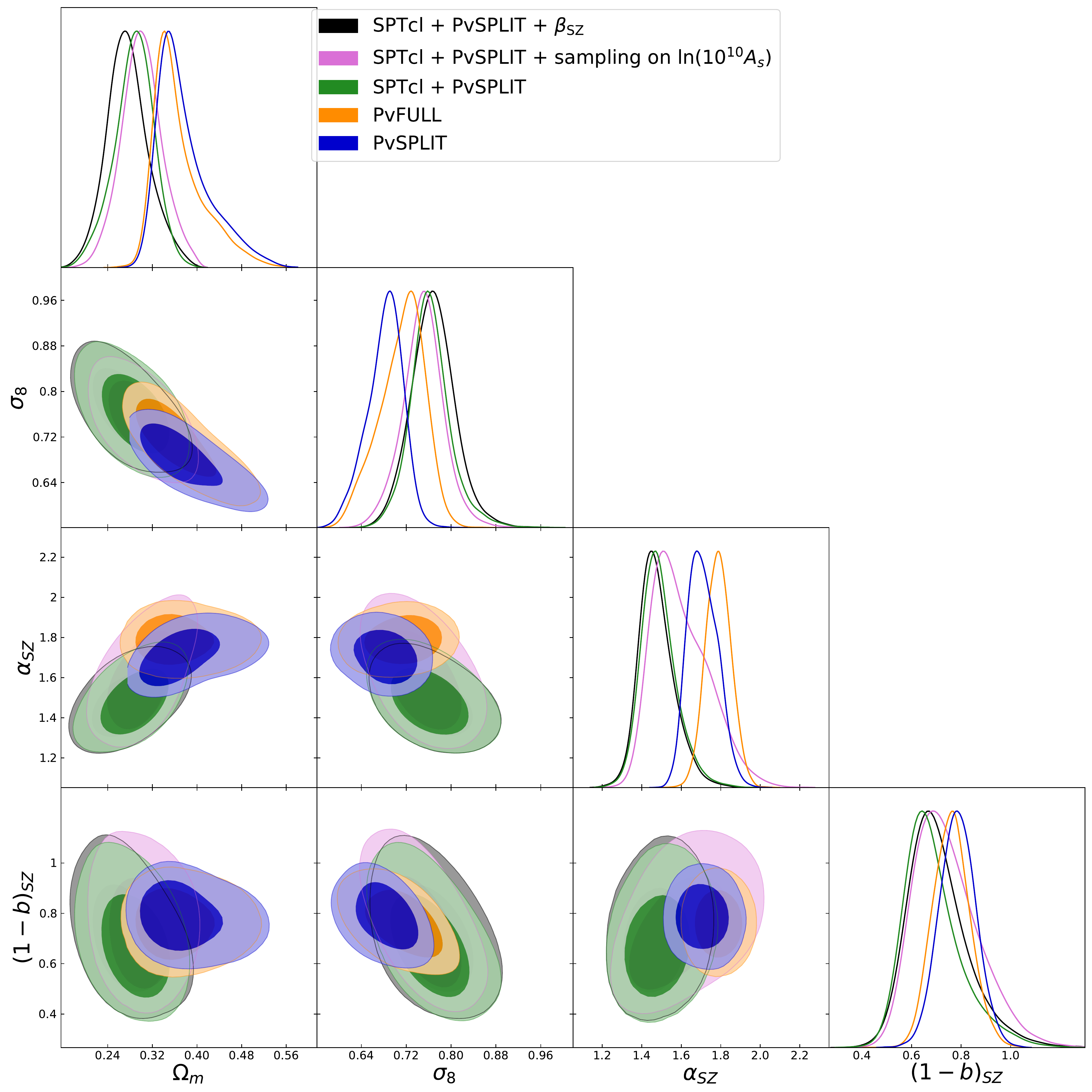}
\caption{We show the one-dimensional and two-dimensional probability distributions for the cosmological ($\Omega_m$, $\sigma_8$) and Planck scaling relation ($\alpha_{\text{SZ}}$, $(1-b)_{\text{SZ}}$) parameters. The contours represent the $68\%$ and $95\%$ CL. We compare results for the original Planck analysis PvFULL (orange contours), with results obtained considering the new Planck likelihood PvSPLIT (dark blue contours). We also show results  for the SPT + Planck combination, comparing the baseline analysis (green contours) with results when considering a logarithmic sampling on $10^{10} A_s$ (pink contours) and when relaxing the assumption of self-similar redshift evolution for Planck scaling relation (black contours). \label{fig:comp_ref1}}
\end{figure*}

The shift we find seems to be due to a combination of different effects. 
First of all, the PvSPLIT likelihood provides slightly different constraints with respect to the original PvFULL one, especially on the $\alpha_{\text{SZ}}$ parameter, already pointing to $ 1.71 _{-0.09}^{+0.07}$, as shown in Fig.~\ref{fig:comp_ref1} (dark blue contours) and in Table~\ref{tab:all_results}.
We then test for the possible impact of sampling choice. 
In particular, as discussed also in \cite{Bocquet:2018ukq}, sampling on $A_s$ or on $\ln {(10^{10}A_s)}$ provides different constraints on the cosmological parameters, where the main effect can be seen on $\Omega_m$ and $H_0$. In our SPTcl + PvSPLIT baseline analysis we are following \cite{Bocquet:2018ukq} and sampling linearly on $A_s$. In the original Planck analysis, the sampling is done on $\ln {(10^{10}A_s)}$, as it is also done for the PvFULL results. We test therefore what happens when considering a logarithmic sampling for the SPTcl + PvSPLIT combination. 
The results are reported in Table~\ref{tab:all_results} and Fig.~\ref{fig:comp_ref1} (pink contours). In this case, for the SPTcl + PvSPLIT + $\ln (10^{10}A_s)$ combination, we find a negligible impact when considering the $\Omega_m$ and $\sigma_8$ constraints. We find a larger effect when focusing on the scaling relation parameters. In particular, the constraints for the mass slope are $ \alpha_{SZ} = 1.60 _{-0.18}^{+0.10}$, being therefore consistent with both the original PvFULL value and the baseline SPTcl + PvSPLIT results.

Nevertheless, the main cause for the departure from self-similarity in the mass slope of the scaling relations is due to the lower value of $\Omega_m$ obtained for the SPTcl + PvSPLIT combination, as it can be seen in Fig.~\ref{fig:baseline_tri}.

As an additional note, we stress that when focusing on the SPT scaling relation parameters (described in Eqs.~\ref{eq:spt_SR1}-\ref{eq:spt_SR3}), results for the SPTcl + PvSPLIT combination are fully consistent with the original analysis presented in \cite{Bocquet:2018ukq}.

As a final test, we relax the assumption of self-similarity for the redshift evolution of the scaling relations, therefore adding $\beta_{\text{SZ}}$ as a varying parameter. 
We report the constraints for the cosmological and scaling relation parameters in Table~\ref{tab:all_results} and Fig.~\ref{fig:comp_ref1} (black contours). We find these results to be fully in agreement with our baseline analysis. For the redshift evolution parameter, we find $\beta_{\text{SZ}} = 0.57 _{-0.51}^{+0.20}$, in agreement with the predicted self-similar value $\beta_{\text{SZ}} = 2/3$.

\subsection{Mass and mass bias evaluation}\label{sec:mass_massbias_results}

We now present the results for the mass and mass bias
evaluation for the clusters in the Planck cosmological sample, following the approach described in section~\ref{sec:me}.

\begin{figure}
\epsscale{1.2}
\plotone{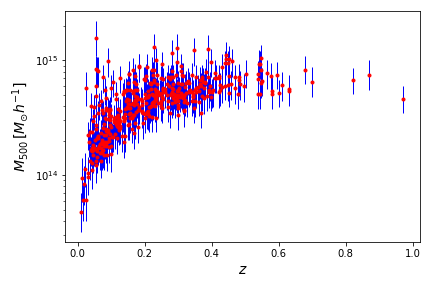}
\caption{Cluster masses for the Planck cosmological sample, evaluated with a Monte-Carlo extraction approach. We show the best-fit value (red points) and the $68\%$ c.l. error bars (in blue). \label{fig:M500_z_MC}}
\end{figure}

\begin{figure*}
\epsscale{1.2}
\plotone{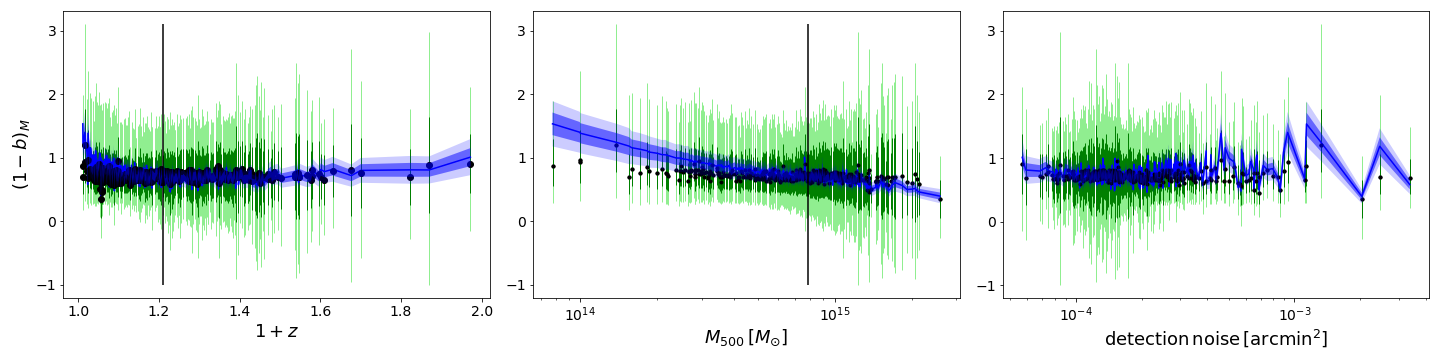}
\caption{Mass bias evaluated from the Monte-Carlo extracted masses $M_{500}$, $(1-b)_M = M_{\text{SZ}}/M_{500}$. We show the mass bias as a function of redshift (left panel), $M_{500}$ (middle panel) and detection noise (right panel): we report the best-fit (black points) with $68\%$ (dark green) and $95\%$ (light green) error bars. The blue shaded area represents the trend and the $68\%$ and $95\%$ CL 
obtained when fitting $M_{\text{SZ}}^{\text{th}}$ from Eq.~\ref{eq:msz_th}, following results in Table~\ref{tab:bias_ev}. The black vertical lines show the values of $z_*$ and $M_*$. \label{fig:bias_Mzn}}
\end{figure*}

In Fig.~\ref{fig:M500_z_MC} we show the results obtained from the Monte-Carlo extraction, presenting the evaluated $M_{500}$ as a function of redshift. These results well reproduce the Planck selection threshold, being able to detect low-mass objects only in the low-redshift regime. 
The full cluster mass catalog is available at \url{https://pole.uchicago.edu/public/data/sptplanck\_cluster}. We report the first entries in Table \ref{tab:masses_full}: in the sixth column we report the constraints on $M_{500}$ and in the seventh column we report the full array of masses extracted through the Monte Carlo approach.
We note that, for the 27 clusters in common with the SPT-SZ catalog, our mass estimation is in agreement within $2 \sigma$ with the estimates from \cite{Bocquet:2018ukq}, as further discussed in Appendix \ref{sec:app2}. The constraints for $(1-b)_M$ are shown in Fig.~\ref{fig:bias_Mzn} in green, with $68\%$ and $95\%$ error bars. Note that the error bars for each cluster are heavily correlated, since they include the marginalization over cosmological and scaling relation parameters starting from the same SPTcl+PvSPLIT baseline chain.

\begin{deluxetable*}{cccccccc}
\tablecaption{First entries for the new Planck cluster catalog. We report cluster ID, coordinates, redshift and signal-to-noise ratio as delivered by Planck collaboration. We add in the sixth and seventh column the evaluation of $M_{500}$ obtained marginalizing over cosmological and scaling relation parameters from our SPTcl+PvSPLIT analysis (labelled as ``free"), and the full array of extracted masses (labelled as ``free,c"). In the eight column we report
the evaluation of $M_{500}$ for the fixed values of cosmological and scaling relation parameters reported in Table \ref{tab:Fixed_ME} (labelled as ``fixed"). The full catalog is available at \url{https://pole.uchicago.edu/public/data/sptplanck\_cluster}. \label{tab:masses_full}}
\tablewidth{10pt}
\startdata
\tablehead{
\colhead{Planck ID$^a$} & \colhead{ra$^a$} & \colhead{dec$^a$} & \colhead{z$^a$} & \colhead{S/N$^a$} &  \colhead{$M_{500}^{\text{free}} [10^{14} M_{\odot} h ^{-1}]$} & \colhead{$M_{500}^{\text{free,c}} [10^{14} M_{\odot} h ^{-1}]$} & \colhead{$M_{500}^{\text{fixed}} [10^{14} M_{\odot} h ^{-1}]$}}
\vspace{0.15cm}
PSZ2 G000.04+45.13 & $229.19051$ & $-1.01722$ & $0.1198$ & $6.75319$ & $3.37 _{-1.11}^{+0.88}$ & $[2.60,...,4.24]$& $3.80 _{-1.04}^{+0.34}$  \\
\vspace{0.15cm}
PSZ2 G000.13+78.04 & $203.55868$ & $ 20.25599$ & $0.171$ & $9.25669$ & $4.52 _{-1.27} ^{+1.11}$ & $[3.40,...,5.38]$& $5.65 _{-0.54}^{+0.91}$\\
\vspace{0.15cm}
PSZ2 G000.40-41.86 & $316.0699$ & $-41.33973$ & $0.1651$ & $8.57995$ & $4.25 _{-1.26} ^{+1.07}$ & $[2.85,...,5.0]$& $5.00 _{-0.83}^{+0.65}$ \\
\vspace{0.15cm}
PSZ2 G000.77-35.69 & $307.97284$ & $-40.59873$ & $0.3416$ & $6.58179$ & $5.31 _{-1.57}^{+1.43}$ & $[3.32,...,3.92]$& $5.81 _{-1.20}^{+0.90}$ \\
\vspace{0.15cm}
PSZ2 G002.77-56.16 & $334.65947$ & $-38.87941$ & $0.1411$ & $9.19606$ & $3.75 _{-1.10}^{+0.92}$ & $[2.90,...,5.52]$& $4.18 _{-1.00}^{+0.29}$\\
\enddata\tablecomments{$^a$from Planck Legacy Archive (\url{https://pla.esac.esa.int)}}
\end{deluxetable*}

We analyze the possible redshift, mass and noise dependence for the mass bias $(1-b)_M$, as defined in Eq.~\ref{eq:bias_Mzn}. The results are obtained from the fit of $M_{\text{SZ}}^{\text{th}} = (1-b)_M^{\text{th}} \cdot M_{500}$ to Planck $M_{\text{SZ}}$ masses, starting again from the SPTcl + PvSPLIT (including $M_{500}$ evaluation) chain. We report these trends in
Fig.~\ref{fig:bias_Mzn} (blue curves)
and the results for the fit in Table~\ref{tab:bias_ev}. 
While we find a value for the amplitude that is consistent with the constraints for $(1-b)_{\text{SZ}}$, having $A_{\text{bias}} = 0.69 _{-0.09}^{+0.04}$, 
we find also strong evidence for mass and redshift evolution. In particular, the mass bias is increasing for high redshift and low mass, with $\gamma_M = -0.41_{-0.06}^{+0.04}$ and $\gamma_z = 0.81\pm 0.13$. Regarding the detection noise, we find no evidence for the mass bias to be dependent on this quantity, since we have $\gamma_n$ consistent with 0 within $1 \, \sigma$.

\begin{deluxetable}{cccc}
\tablecaption{Parameters for the mass, redshift and detection noise dependence of the mass bias obtained when fitting Eq.~\ref{eq:msz_th}}. We report the $68\%$ CL constraints.\label{tab:bias_ev}
\tablewidth{10pt}
\tablehead{
\colhead{Parameter} & \colhead{Value} & \colhead{Parameter} & \colhead{Value}}
\vspace{0.15cm}
\startdata
\vspace{0.15cm}
$A_{\text{bias}}$ & $0.69 _{-0.09}^{+0.04}$ & $\gamma_z$ & $ 0.81\pm 0.13$ \\
\vspace{0.15cm}
$\gamma_M$ & $ -0.41_{-0.06}^{+0.04}$ & $\gamma_n$ & $ 0.05 _{-0.08}^{+0.06}$ \\
\enddata
\end{deluxetable}

We conclude this section presenting masses for the PSZ2 cosmo catalog obtained when fixing the cosmological and scaling relation parameters.
For the cosmological parameters, we adopt a flat $\nu \Lambda$CDM scenario, following \cite{Bocquet:2018ukq}. For the Planck scaling relation parameters, we take the best-fit values from the SPTcl + PvSPLIT baseline run with the fixed cosmology. The values of the parameters are reported in Table~\ref{tab:Fixed_ME}. Also in this case, the full cluster mass catalog is available at \url{https://pole.uchicago.edu/public/data/sptplanck\_cluster}. We report the first entries in Table \ref{tab:masses_full}, eighth column. As for the marginalized masses, for the 27 clusters in common with the SPT-SZ catalog, our mass estimation is in agreement within $2 \sigma$ with the estimates from \cite{Bocquet:2018ukq}, as further discussed in Appendix \ref{sec:app2}.

\begin{deluxetable}{cccc}
\tablecaption{Fixed values of cosmological and Planck scaling relation parameters used to evaluate cluster masses. \label{tab:Fixed_ME}}
\tablewidth{10pt}
\startdata
\tablehead{
\colhead{Parameter} & \colhead{Value} & \colhead{Parameter} & \colhead{Value}}
$\Omega_m$ & $0.3$ & $\alpha_{\text{SZ}}$ & $1.62$ \\
$\sigma_8$ & $ 0.8$ & $\beta_{\text{SZ}}$ & $0.67 $ \\
$\Omega_{\nu}h^2$ & $ 0.00064 $ & $\sigma_{\ln{Y},\text{SZ}}$ & $0.07 $  \\
$\log{Y_*}$ & $-0.14 $ & $(1-b)_{\text{SZ}}$ & $0.58 $\\
\enddata
\end{deluxetable}

\begin{figure}
\epsscale{1.2}
\plotone{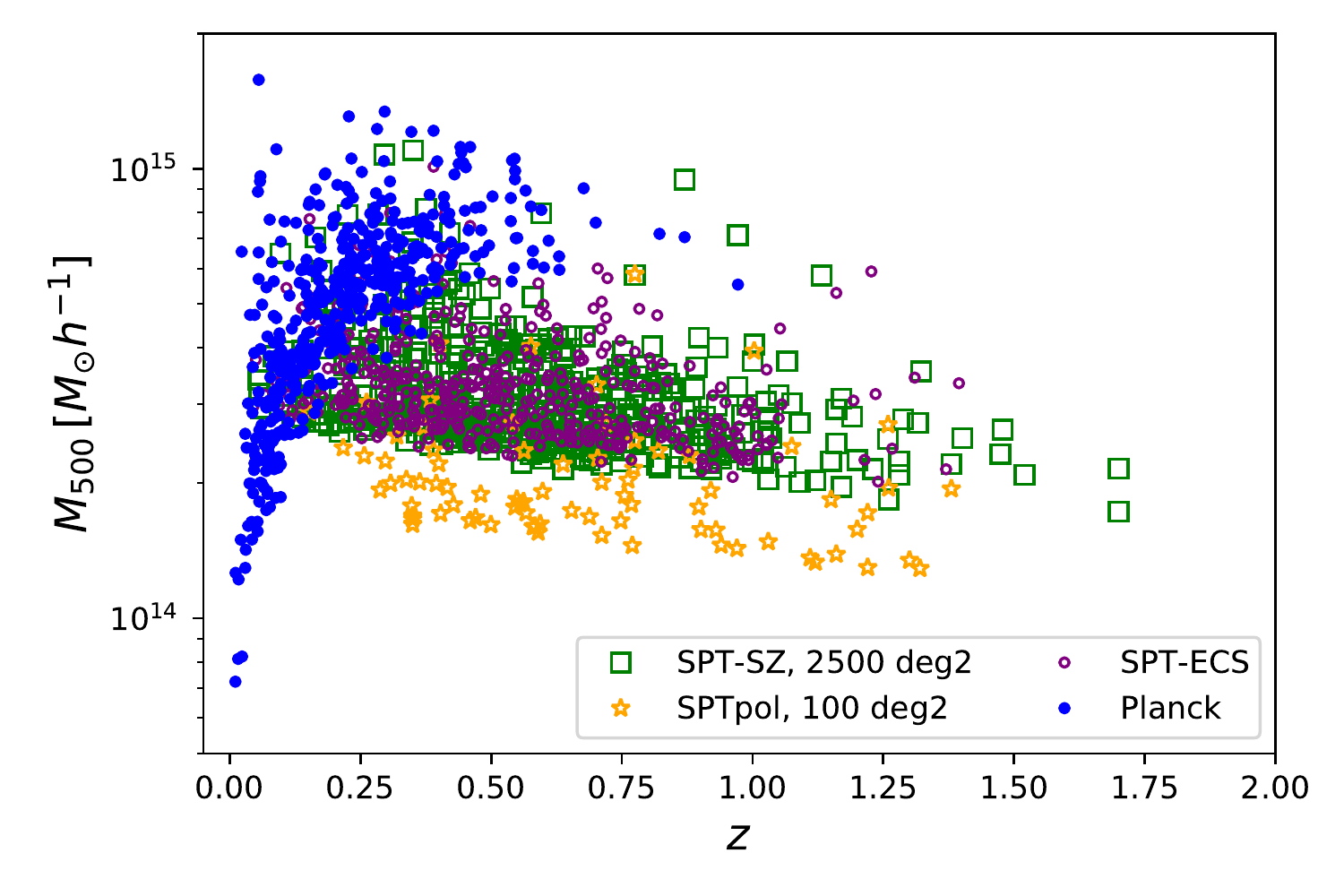}
\caption{Evaluation of cluster masses, for each cluster in the Planck cosmological cluster sample (blue points), for fixed values of cosmological and scaling relation parameters. We report also the SPT cluster masses from the SPT-SZ 2500 deg$^2$ catalog (green squares), from the SPTpol 100 deg$^2$ catalog (yellow stars) and from the SPTpol Extended cluster catalog (purple circles). \label{fig:Mass_fixed}}
\end{figure}

We show in Fig.~\ref{fig:Mass_fixed} the Planck evaluated masses $M_{500}$, as function of redshift, in comparison with cluster masses from the SPT-SZ 2500 deg$^2$ catalog \citep{Bocquet:2018ukq}.
As a reference, we also add clusters from recent SPT observations: the 79 clusters from the SPTpol 100 deg$^2$ sample \citep{Huang:2019qnh}, and the 448 clusters from the SPTpol Extended (SPT-ECS) sample \citep{Bleem:2019fvp}.

\section{Discussion}\label{sec:discussion}

The results presented in the previous sections show the tight correlation between cosmological and scaling relation parameters, highlighting that a correct and unbiased evaluation of cluster masses is fundamental to perform precision cosmology with galaxy clusters. 
In an ideal scenario, to calibrate the scaling relations we would rely on high-precision multi-wavelength observations for each cluster in the considered cosmological sample. Since current counterpart observations in X-rays and optical bands do not cover the full Planck cosmological cluster sample, in this analysis we choose an alternative approach, by exploiting the cosmological constraining power of SPT-SZ cluster catalog, with its internal X-ray and WL mass calibration, and use the Planck-SPT combination to constrain the Planck scaling relations.

The results presented in section~\ref{sec:results} point towards the necessity of improving the general astrophysical model adopted for the cluster evolution.
We start discussing the results obtained for the SPTcl + PvSPLIT cluster catalog combination.
First of all, we highlight the powerful cosmological constraining power of the SPT-SZ cluster sample: SPT data are driving the results, pushing the constraints for the SPTcl + PvSPLIT combination.
For this dataset combination, we are also able to get tight constraints on the Planck scaling relation parameters, comparable with the results from PvFULL (i.e. the original full Planck likelihood), as shown in Table~\ref{tab:all_results}. In particular, we decide to focus on the parameters describing the mass dependence, therefore not considering external calibration and assumption of self-similarity for the mass bias, described by $(1-b)_{\text{SZ}}$, and the mass slope $\alpha_{\text{SZ}}$. 

For the mass bias, we find $(1-b)_{\text{SZ}} = 0.69 _{-0.14}^{+0.07}$. This is still in agreement within $2 \, \sigma$ with the different external WL calibrations and hydro-dynamical simulation estimations, but it also encompasses the lower values preferred from CMB data. This result can be further discussed in light of the evaluation of $(1-b)_M$ that we performed for each single cluster. We discuss in section~\ref{sec:mass_bias} the difference between the scaling relation parameter and the measured mass bias. The two quantities describe from different approaches a general non-precise knowledge of how the astrophysical processes affect the theoretical model for the cluster evolution (and as a consequence how we model the mass-observable relation and the selection approach). 
By analyzing $(1-b)_M$, we find strong hints for mass and reshift evolution of this quantity, with the amplitude being consistent with $(1-b)_{\text{SZ}}$, having $A_{\text{bias}} = 0.69 _{-0.09}^{+0.04}$, as shown in Table~\ref{tab:bias_ev}. The increasing trend for the redshift evolution is also consistent with the analysis shown in \cite{Salvati:2019zdp}.

We now focus on the mass slope of the scaling relations, $\alpha_{\text{SZ}}$. For SPTcl + PvSPLIT we find $\alpha_{\text{SZ}} = 1.49 _{-0.10}^{+0.07}$, which is $\sim 4 \, \sigma$ lower than the self-similarity value. As discussed in section~\ref{sec:results_cosmo}, this low value is due to a combination of different effects, with the dominant one being the shift of $\Omega_m$ towards lower values. Indeed, this shift slightly tilts the mass function, such that it leads to fewer objects at low mass and more objects in the high-mass tail. The low value of $\alpha_{\text{SZ}}$ seems to accommodate for this tilt, balancing the low-mass high-mass weight. 
The mass-redshift evolution of $(1-b)_M$ seems to account for the same effect, balancing the low-mass high-mass trend.
We also stress that, when not assuming self-similarity for the redshift evolution of Planck scaling relation and sampling also on the $\beta_{\text{SZ}}$ parameter, we find consistent results with the baseline analysis and no evidence for departure from self-similarity.

From these combined results on the Planck scaling relation parameters and the estimated mass bias,
we can take one main message: the simple model for the mass calibration of tSZ clusters, based on the assumptions of self-similarity, spherical symmetry and hydrostatic equilibrium, needs to be improved towards a more realistic description, at least for the modelling of the mass (and therefore scale) dependence.
This is indeed the approach used for the SPT-SZ cluster analysis: the empirical, multi-observable approach used for the mass calibration provides constraints for the different parameters (defined in Eqs.\ref{eq:spt_SR1}-\ref{eq:spt_SR3}) not relying on strong theoretical assumptions. 

As a last point, we discuss the dependence of the measured mass bias with respect to the detection noise. As described in section \ref{sec:mass_bias}, with this parametrization we try to quantify the impact of the detection process in the full cosmological modelling. From our analysis, we find no hint for a noise dependence of the mass bias, having $\gamma_n = 0.05 _{-0.08}^{+0.06}$. As a further test, we check the results when considering only the noise dependence for the bias, i.e.
\begin{equation}
    (1-b)_M^{\text{th}} = A_n \left( \dfrac{\sigma_{\text{f}}(\theta_{500},l,b)}{\sigma_{\text{f},*}(\theta_{500})}\right)^{\gamma_n} \, .
\end{equation}
In this case, we find $A_n = 0.60 _{-0.14}^{+0.06}$ and $\gamma_n = -0.37 _{-0.12}^{+0.14}$, pointing to a decreasing trend of the measured bias with respect to the noise. This implies that the $M_{\text{SZ}}$ estimation for clusters detected in patches with higher detection noise is more biased, possibly due to a loss of tSZ signal. 

On the other hand, when considering only the mass and redshift dependence for the measured mass bias, we find results for the amplitude and the slopes that are fully consistent with what we report in Tab.~\ref{tab:bias_ev}. This stresses even more that an incorrect characterization of the mass and redshift dependence for the mass-observable relation is still a dominant source of uncertainties with respect to possible systematics coming from the modelling of the cluster selection process.

\section{Conclusion}\label{sec:conclusion}

In this paper we provide the first combination of Planck and SPT cluster catalogs for a cosmological analysis, with the aim of exploiting the SPT cosmological constraining power to provide an independent evaluation of Planck scaling relation parameters. We build a new likelihood (labelled ``PvSPLIT") to analyze the Planck PSZ2 cosmo sample, removing the clusters and sky patches in common with SPT observations.

The baseline analysis is given by the ``SPTcl + PvSPLIT" combination, where we do 
not 
rely on the
external X-ray and WL calibrations for the mass slope $\alpha_{\text{SZ}}$ and the mass bias $(1-b)_{\text{SZ}}$ 
adopted in the original Planck analysis.

We summarize our main findings below:
\begin{enumerate}

\item We show the strong constraining power of SPT-SZ clusters, which drives the results for the SPTcl + PvSPLIT combination. 
Focusing on Planck scaling relation parameters, we find that the SPTcl + PvSPLIT combination provides results comparable in accuracy with the external X-ray and WL calibrations used for the original Planck analysis, having $\alpha_{\text{SZ}} = 1.49 _{-0.10}^{+0.07}$ and $(1-b)_{\text{SZ}} = 0.69 _{-0.14}^{+0.07}$. We stress that the value of $\alpha_{\text{SZ}}$ that we find is $\sim 4 \, \sigma$ 
lower than the expected self-similar value, $\alpha_\text{SZ} = 1.8$, a result driven primarily by the relatively low values of $\Omega_m$ preferred from SPT data.

\item Through a Monte Carlo extraction approach, we provide new estimates of Planck cluster masses $M_{500}$, obtained marginalizing over 
cosmological and scaling relation parameter posteriors derived from the SPTcl + PvSPLIT analysis.
We provide also an evaluation of $M_{500}$ masses for Planck clusters in the PSZ2 cosmo catalog at fixed values of cosmological and scaling relation parameters. The cluster mass catalogs are available at \url{https://pole.uchicago.edu/public/data/sptplanck\_cluster}.

\item We provide a measurement of the mass bias, $(1-b)_M$, for 433 over 439 clusters of the PSZ2 cosmo sample 
(for which we have the redshift in the original Planck analysis), using the $M_{\text{SZ}}$ measurements from Planck and our estimation of $M_{500}$. The constraints for $(1-b)_M$ 
account for the uncertainties on the cosmological and scaling relation parameters derived in this work.

We study a possible dependence of $(1-b)_M$ with respect to the cluster mass and redshift, and to the survey detection noise. The aim is to highlight the impact, in the cosmological analysis, of the assumed modelling for the mass-observable relation and the cluster detection approach. On the one hand, we find 
$(1-b)_M$ to have a decreasing trend with respect to the cluster mass and an increasing trend with respect to redshift, with the slopes being $\gamma_M = -0.41_{-0.06}^{+0.04}$ and $\gamma_z = 0.81\pm 0.13$.
On the other hand, we do not see any noise dependence, having $\gamma_n$ fully consistent with 0. 

\item Comparing the results for the scaling relation parameters and the measured mass bias dependencies, we find them to mimic the same effects, mainly a departure from self-similarity for the cluster evolution, and therefore the necessity to consider different dependencies for the low-mass vs. high-mass and low-redshift vs. high-redshift clusters.

\end{enumerate}

This analysis confirms the importance of an accurate mass calibration when using 
cluster counts as a cosmological probe.
We find that the simple model for the mass calibration  of  tSZ  clusters,  based  on  the  assumptions  of self-similarity, spherical symmetry and hydrostatic equilibrium, needs to be improved towards a more realistic description. Furthermore, we stress that the adopted modelling should take into the cluster sample selection, from the cluster mass-redshift distribution to the impact of the detection approach. 
This project is paving the way towards a full joint analysis of SPT and Planck cluster catalogs, with a joint mass-calibration, allowing to more stringent tests of cosmology beyond flat $\nu \Lambda$CDM scenario.

\begin{acknowledgments}

The authors thank the anonymous referee for the useful comments they provided to improve the presentation and discussion of the analysis.
LS and MC are supported by ERC-StG "ClustersXCosmo" grant agreement 716762. 
AS is supported by the ERC-StG ‘ClustersXCosmo’ grant agreement 716762 and by the FARE-MIUR grant ‘ClustersXEuclid’ R165SBKTMA and INFN InDark Grant.
TS acknowledges support from the German Federal Ministry for Economic Affairs und Energy (BMWi) provided through DLR under projects 50OR2002 and 50OR2106, as well as support provided by the Deutsche Forschungsgemeinschaft (DFG, German Research Foundation) under grant 415537506.
The Melbourne group acknowledges support from the Australian Research Council’s Discovery Projects scheme (DP200101068).

This research made use of: computation facilities of CINECA, within the projects INA17\_C5B32, INA20\_C6B51, INA21\_C8B43, and at the Observatory of Trieste \citep{2020ASPC..527..307T,2020ASPC..527..303B};
observations obtained with Planck (\url{http://www.esa.int/ Planck}), an ESA science mission with instruments and contributions directly funded by ESA Member States, NASA, and Canada; the SZ-Cluster Database (\url{http://szcluster-db.ias.u-psud.fr}) operated by the Integrated Data and Operation Centre (IDOC) at the Institut d’Astrophysique Spatiale (IAS) under contract with CNES and CNRS.

The South Pole Telescope program is supported by the National Science Foundation (NSF) through grants PLR-1248097 and OPP-1852617. Partial support is also provided by the NSF Physics Frontier Center grant PHY- 1125897 to the Kavli Institute of Cosmological Physics at the University of Chicago, the Kavli Foun- dation, and the Gordon and Betty Moore Foundation through grant GBMF$\#947$ to the University of Chicago. Argonne National Laboratory’s work was supported by the U.S. Department of Energy, Office of Science, Of- fice of High Energy Physics, under contract DE-AC02- 06CH11357.
\textit{Facilities}: Magellan: Clay (Megacam),
Hubble Space Telescope, Chandra, Gemini:South (GMOS), Magellan: Clay (PISCO).
\end{acknowledgments}

\appendix

\section{Further analysis}\label{sec:app1}

\begin{figure*}
\epsscale{0.85}
\plotone{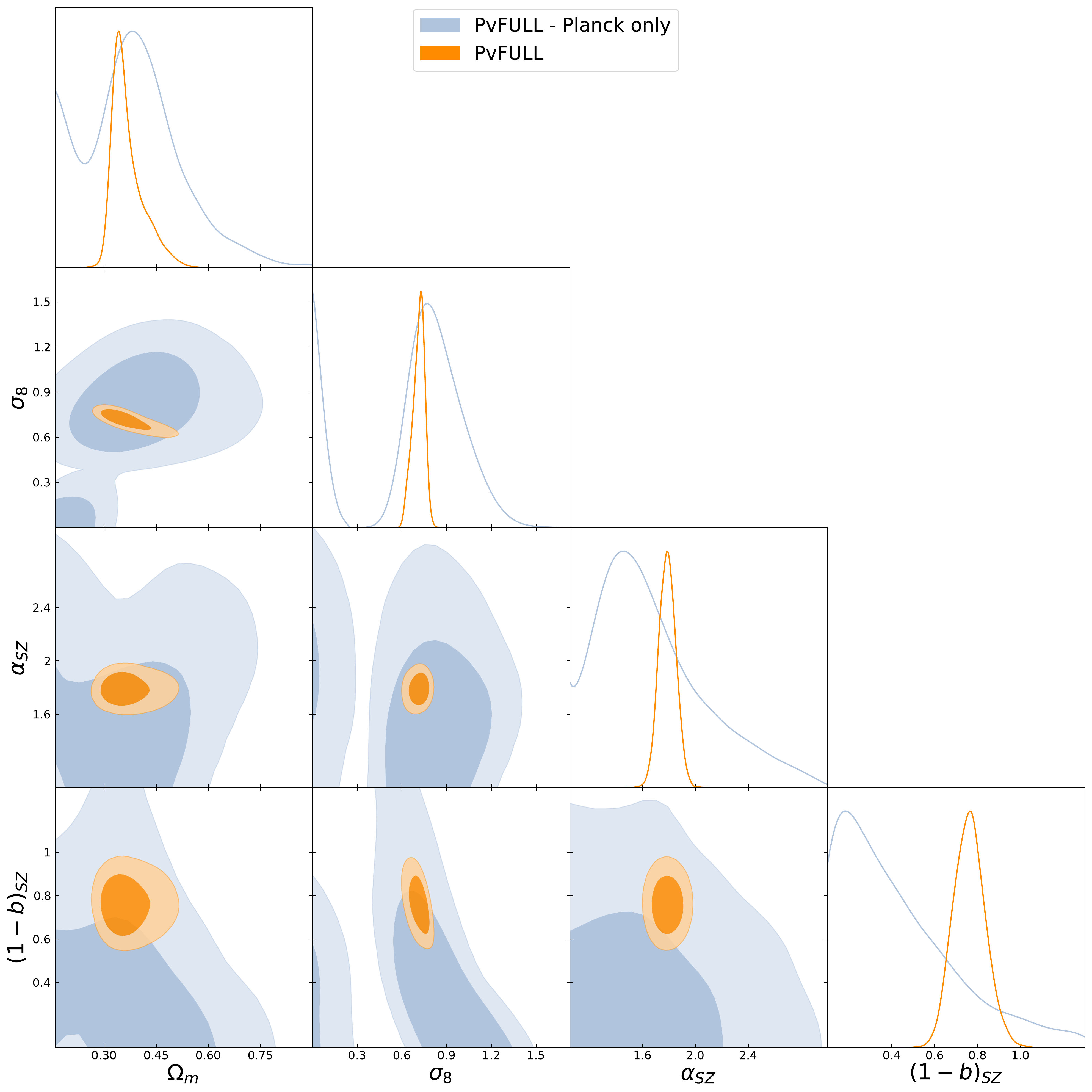}
\caption{We show the one-dimensional and two-dimensional probability distributions for the cosmological ($\Omega_m$, $\sigma_8$) and Planck scaling relation ($\alpha_{\text{SZ}}$, $(1-b)_{\text{SZ}}$) parameters. The contours represent the $68\%$ and $95\%$ CL. We compare results for the complete Planck analysis (Planck clusters + BAO + BBN + external scaling relation calibrations assuming self-similar redshift evolution) in orange, with results obtained considering only Planck clusters (light blue).\label{fig:PvFULL_comp}}
\end{figure*}

We show here results for different analyses of Planck cluster data in the $\nu \Lambda$CDM scenario, in order to further discuss the constraints presented in section \ref{sec:results_cosmo}. In Fig.~\ref{fig:PvFULL_comp} we report the results for the original Planck analysis, PvFULL, obtained from the combination of the Planck cluster cosmological catalog, with external X-ray and WL calibrations, BAO data  \citep{Alam:2016hwk,Beutler:2012px,10.1111/j.1365-2966.2011.19250.x,Ross:2014qpa} and BBN constraints on the baryon density.
As a comparison, we show the results obtained from Planck cluster catalog without external datasets and X-ray and WL calibrations on the mass bias and mass slope of the scaling relations, following the approach used when building the SPTcl + PvSPLIT likelihood. As expected, removing information from mass calibration and external datasets largely reduce the constraining power of galaxy clusters. These results confirm the strength of combining Planck and SPT cluster catalogs in providing constraints for the full cosmological and scaling relation parameter space.

\begin{figure*}
\epsscale{0.85}
\plotone{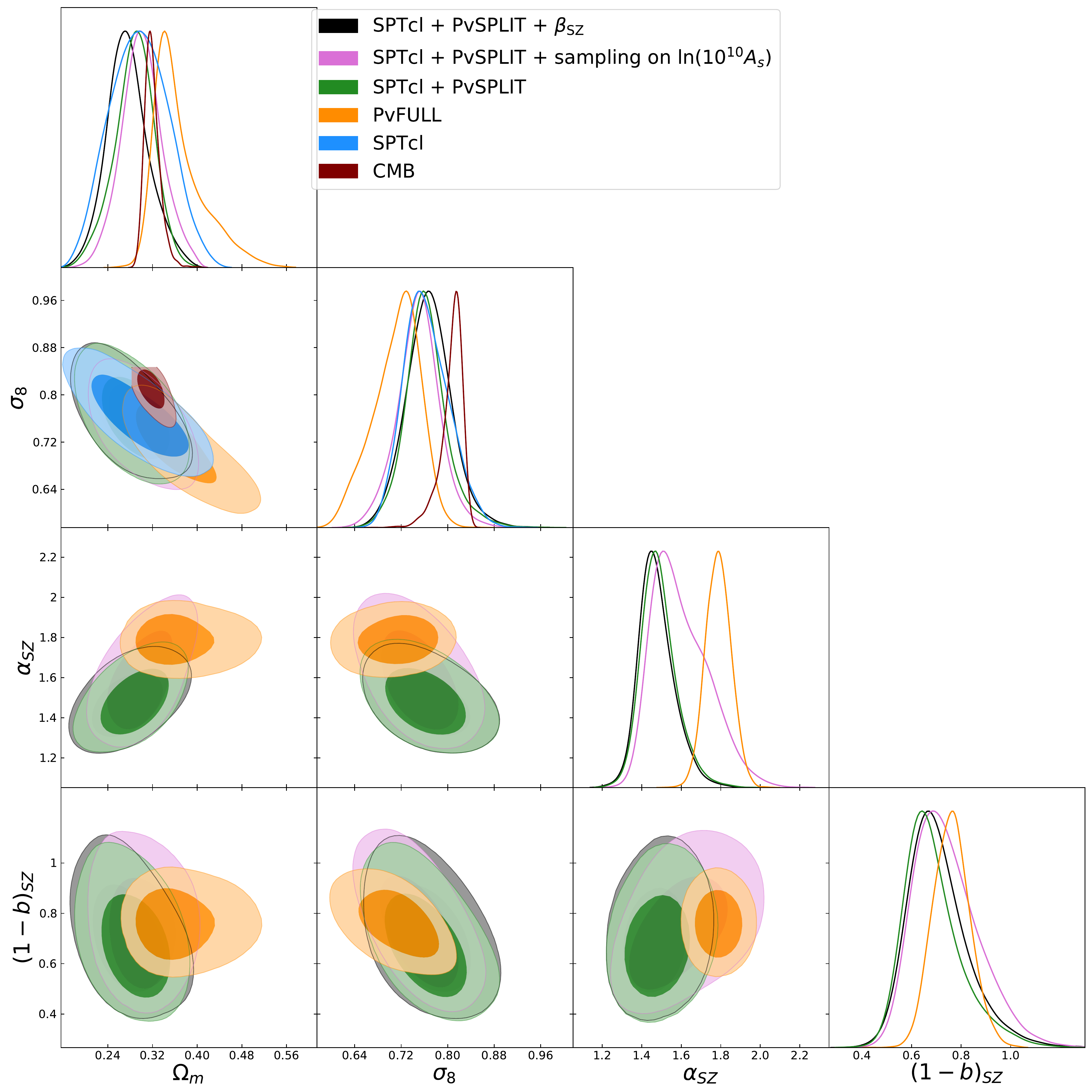}
\caption{We show the one-dimensional and two-dimensional probability distributions for the cosmological ($\Omega_m$, $\sigma_8$) and Planck scaling relation ($\alpha_{\text{SZ}}$, $(1-b)_{\text{SZ}}$) parameters. The contours represent the $68\%$ and $95\%$ CL. We compare results for different dataset combinations, as described in the text.\label{fig:all_comb}}
\end{figure*}

We conclude this section showing in Fig.~\ref{fig:all_comb} a full comparison of the different dataset combinations introduced in section~\ref{sec:results_cosmo}. 
We also show the constraints from the latest Planck CMB analysis \citep{2020A&A...641A...6P} (brown contours). The MCMC chains are taken from the Planck Legacy Archive\footnote{\url{https://pla.esac.esa.int}}.

\section{Mass comparison}\label{sec:app2}

\begin{figure*}
\epsscale{0.85}
\plotone{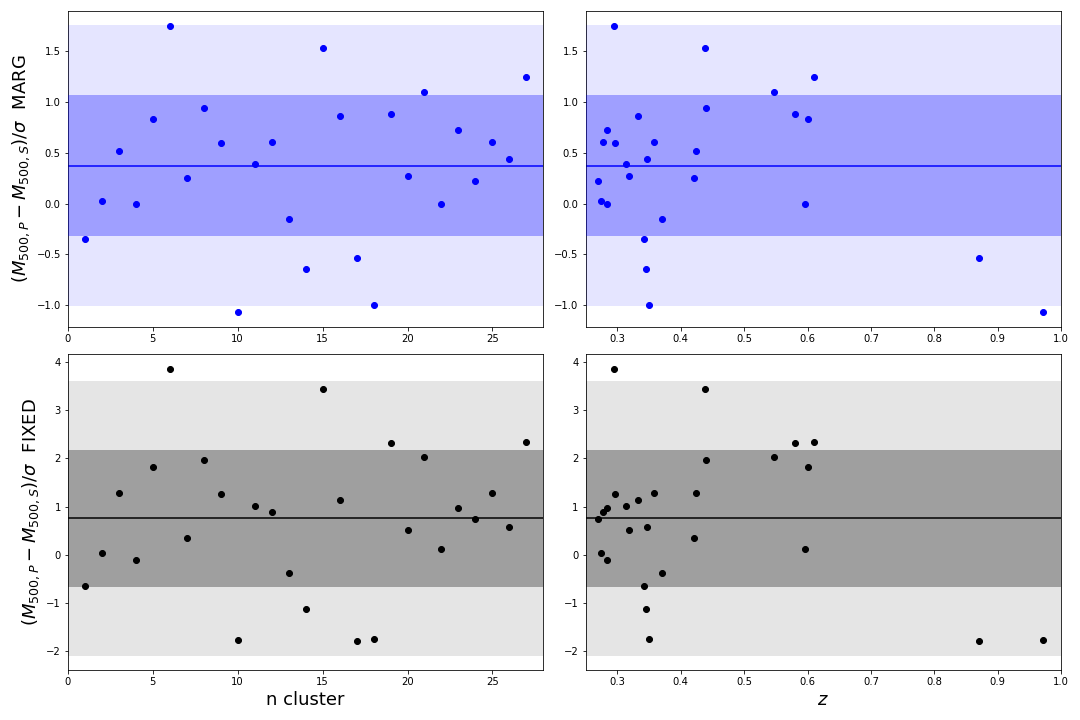}
\caption{We show the distribution of mass differences for the 27 clusters in common between Planck and SPT-SZ cosmological catalogs, as a function of the cluster number (left) and redshift (right). We compare the results when considering the mass estimates marginalized over cosmological and scaling relation parameters (in blue, top panels) and obtained with fixed cosmological and scaling relation parameters (in black, bottom panels). The shaded areas represent the $1 \sigma$ and $2 \sigma$ intervals of the distribution. \label{fig:mass_comp_PSPT}}
\end{figure*}

We compare the cluster masses for the 27 clusters in common between Planck and SPT-SZ cosmological catalog, for redshift $z > 0.25$. For the SPT-SZ masses, we consider estimates from \cite{Bocquet:2018ukq} available on the SPT webpage\footnote{ \url{https://pole.uchicago.edu/public/data/sptsz-clusters/}}. For the Planck masses, we make use of our Monte-Carlo estimates. 
In Fig.~\ref{fig:mass_comp_PSPT} we show the quantity $\Delta_M = (M_{500,P}-M_{500,S})/\sigma$,
where we define $\sigma = \sqrt{\sigma^2_{\text{Planck}} + \sigma^2_{\text{SPT}}}$.
We consider the mass estimates obtained marginalizing over cosmological and scaling relation parameters (in blue, top panels) and fixing the parameters (in black, bottom panels). We note that the values are quite spread, nevertheless showing a consistency within $2 \sigma$ between the different estimations. The agreement is stronger for the marginalized estimates, $\Delta_M = 0.37 \pm 0.69$, than for estimations at fixed cosmological and scaling relation parameters, $\Delta_M = 1.25 \pm 1.41$.  

\begin{figure*}
\epsscale{0.85}
\plotone{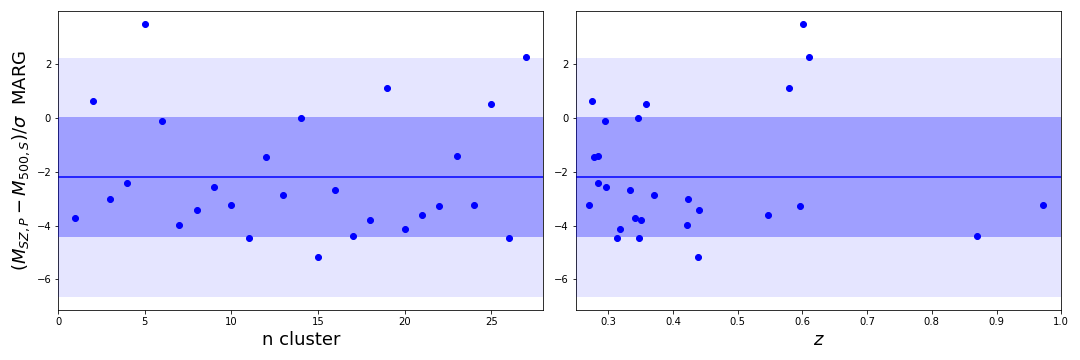}
\caption{We show the distribution of mass differences for the 27 clusters in common between Planck and SPT-SZ cosmological catalogs, as a function of the cluster number (left) and redshift (right). 
For the Planck clusters, we consider $M_{\text{SZ}}$ estimates. For the SPT-SZ clusters we consider the mass estimates marginalized over cosmological and scaling relation parameters. The shaded areas represent the $1 \sigma$ and $2 \sigma$ intervals of the distribution. \label{fig:mass_comp_PszSPT}}
\end{figure*}

As a further test, we show the comparison between Planck $M_{\text{SZ}}$ masses from \cite{2016A&A...594A..24P} and the SPT marginalized masses, in Fig.~\ref{fig:mass_comp_PszSPT}. In this case we find $\Delta_M = -2.20 \pm 2.22$, clearly showing that $M_{\text{SZ}}$ estimations are biased low.

\bibliography{sample63}{}
\bibliographystyle{aasjournal}

\end{document}